\def \ha {H{$_\alpha$}}
\def \pv {P{\sc v}}
\def \vinf {$\varv_\infty$}
\def \kms {$\rm km\,s^{-1}$}
\def \mdotun {$10^{-6}\,{\rm M_{\odot}/yr}$}   
\def \mdot {$\dot M$}

\def \rstar {$R_\star$}
\def \rstun  {$R_\odot$}

\def \teff {$T_{\rm eff}$}
\def \logg {$\log g$}

\def \fcl   {$f_{\rm cl}$}
\def \xic   {$x_{\rm ic}$}
\def \dt    {$\delta t$}
\def \dv    {$\delta \varv$}
\def \dvb   {$\delta \varv_\beta$}
\def \fv    {$f_{\rm v}$}
\def \fvel  {$f_{\rm vel}$}

\def \Dv     {$\Delta \varv$} 

\def \one   {Paper\,I}
\def \lcep  {$\lambda$~Cep}
\def \zpup  {$\zeta$~Pup}
\def \taucl {$\tau_{\rm cl}$}
\def \wl    {$W_\lambda$}
\def \heii  {He\,{\sc ii}\,4686}

\documentclass{aa}  
\usepackage{graphicx}
\usepackage{txfonts}
\usepackage{natbib}
\usepackage{url}

\begin{document}
   \title{Mass loss from inhomogeneous hot star winds}

   \subtitle{II. Constraints from a combined optical/UV study}
   \author{J.O. Sundqvist\inst{1}\and
           J. Puls\inst{1}\and 
           A. Feldmeier\inst{2}\and
           S.P. Owocki\inst{3}
           }
          
   \institute{Universit\"atssternwarte M\"unchen, Scheinerstr. 1, 81679 M\"unchen, Germany\\
              \email{jon@usm.uni-muenchen.de}\and Institut f\"ur Physik und Astronomie,
              Karl-Liebknecht-Strasse 24/25, 14476 Potsdam-Golm, Germany\and 
              University of Delaware, Bartol Research Institute, Newark, DE 19716, USA}
   \date{Received 2010-09-16 / Accepted 2011-01-18}

 \abstract
       { Mass loss is essential for massive
       star evolution, thus also for the variety of
       astrophysical applications relying on its predictions. However,
       mass-loss rates currently in use for hot, massive stars have
       recently been seriously questioned, mainly because of the
       effects of \textit{wind clumping}. }
       { We investigate the impact of clumping on diagnostic
         ultraviolet resonance and optical recombination lines often
         used to derive empirical mass-loss rates of hot stars.
         Optically thick clumps, a non-void interclump medium, and a
         non-monotonic velocity field are all accounted for in a
         single model. The line formation is first theoretically
         studied, after which an exemplary multi-diagnostic study of
         an O-supergiant is performed. }
       { We used 2D and 3D stochastic and radiation-hydrodynamic wind
       models, constructed by assembling 1D snapshots in radially
       independent slices. To compute synthetic spectra, we developed
       and used detailed radiative transfer codes for both
       recombination lines (solving the `formal integral') and
       resonance lines (using a Monte-Carlo approach).  In addition,
       we propose an analytic method to model these lines in clumpy
       winds, which does not rely on optically thin clumping. }
      { The importance of the `vorosity' effect for line formation in
       clumpy winds is emphasized. Resonance lines are generally more
       affected by optically thick clumping than recombination
       lines. Synthetic spectra calculated directly from current
       radiation-hydrodynamic wind models of the line-driven
       instability are unable to in parallel reproduce strategic
       optical and ultraviolet lines for the Galactic
       O-supergiant \lcep. Using our stochastic wind models, we obtain
       consistent fits essentially by increasing the clumping in the
       inner wind. A mass-loss rate is derived that is approximately
       two times lower than what is predicted by the line-driven wind
       theory, but much higher than the corresponding rate derived
       when assuming optically thin clumps. Our analytic formulation
       for line formation is used to demonstrate the potential
       importance of optically thick clumping in diagnostic lines in
       so-called weak-winded stars and to confirm recent results that
       resonance doublets may be used as tracers of wind structure and
       optically thick clumping. }
       { We confirm earlier results that a re-investigation of the
       structures in the inner wind predicted by line-driven
       instability simulations is needed. Our derived mass-loss rate
       for \lcep~suggests that only moderate reductions of current
       mass-loss predictions for OB-stars are necessary, but this
       nevertheless prompts investigations on feedback effects from
       optically thick clumping on the steady-state, NLTE wind models used
       for quantitative spectroscopy. }


   \maketitle
%

\section{Introduction}
\label{Introduction}

Massive stars are fundamental in many fields of modern astrophysics.
In the present Universe, they dynamically and chemically shape their
surroundings and the interstellar medium by their output of ionizing
radiation, energy and momentum, and nuclear processed material. In the
distant Universe, they dominate the ultraviolet (UV) light from young
galaxies. Indeed, massive stars may be regarded as `cosmic
engines' \citep{Bresolin08}.  Hot, massive stars possess strong and
powerful winds that affect evolutionary time scales, chemical surface
abundances, and luminosities. In fact, changing the mass-loss rates of
massive stars by only a factor of two has a dramatic effect on their
overall evolution \citep{Meynet94}.  The winds from these stars are
described by the radiative line-driven wind theory, in which the
standard model \citep[based on the pioneering works
by][]{Lucy70,Castor75} assumes that the wind is stationary, spherically
symmetric, and homogeneous. Despite this theory's apparent
success \citep[e.g.,][]{Vink00}, theoretical as well as observational
evidence of an inhomogeneous, time-dependent wind has become overwhelming 
in the past years \citep[for a comprehensive summary,
see][]{Puls08}.

Direct simulations of the time-dependent wind have confirmed that the
so-called line-driven instability causes a highly structured wind in
both density and velocity \citep{Owocki88,Feldmeier95,Dessart05}.
Much indirect evidence of such \textit{small-scale inhomogeneities}
(clumping) has arisen from quantitative spectroscopy. Clumping has
severe consequences for any interpretation of observed spectra, with
the inferred mass-loss rates particularly affected. When deriving
mass-loss rates from observations, wind clumping has traditionally
been accounted for by assuming \textit{optically thin} clumps and a
void interclump medium, while keeping a smooth velocity field.
Results based on this \textit{microclumping} approach have, for
example, led to a downward revision of empirical mass-loss rates from
Wolf-Rayet (WR) stars by roughly a factor of three
\citep[reviewed in][]{Crowther07}.  

However, for O stars, highly clumped winds with very low mass-loss
rates must be invoked in order to reconcile investigations of
different diagnostics within the microclumping model. The most
alarming example was the phosphorus {\sc v} (\pv) UV analysis
by \citet{Fullerton06}, which indicated reductions of previously
accepted values by an order of magnitude (or even more), with dwarfs,
giants, and supergiants all affected (but see
also \citeauthor{Waldron10}~2010, who argued that {\sc xuv} radiation
could seriously alter the ionization fractions of \pv). Such low
mass-loss rates would be in stark contrast with the predictions of
line-driven wind theory and have dramatic consequences for the
evolution of, and feedback from, massive stars. Naturally, the widely
discrepant values inferred from different observations and diagnostics
drastically lower the reliability of mass-loss rates currently in use,
and an explanation is urgently needed. A key question is whether the
microclumping model fails to deliver accurate empirical rates under
certain conditions.

Simplified techniques to account for optically thick clumps in X-ray
line formation have been developed \citep{Feldmeier03,Owocki04}, but
it has yet to be settled whether or not this is important to consider
when deriving empirical mass-loss rates from these
diagnostics \citep{Oskinova06,Cohen10}. First attempts to relax the
assumptions of the microclumping model for UV resonance lines were
made by \citet{Oskinova07} (optically thick clumps), \citet{Zsargo08}
(a non-void interclump medium), and
\citet{Owocki08} (a non-monotonic velocity field). \citet{Sundqvist10a}
(hereafter Paper\,I) carried out the first detailed investigation,
relaxing \textit{all} the above assumptions, and showed that, indeed,
the microclumping approximation is not a suitable assumption for UV
resonance line formation under conditions prevailing in typical
OB-star winds. Recently, these results were empirically supported for
the case of B supergiants by \citet{Prinja10}, who analyzed
profile-strength ratios of the individual components of resonance line
doublets and found that the observed ratios were inconsistent with
lines formed in a smooth or `microclumped' wind. Furthermore, Paper\,I
demonstrated that resonance line profiles calculated from 2D,
stochastic wind models were compatible with mass-loss rates an order
of magnitude higher than those derived from the same lines but using
the microclumping technique. However, as pointed out in that paper, a
consistent modeling of the resonance lines also introduces
degeneracies among the parameters used to define the wind structure,
degeneracies that can only be broken by considering different
diagnostics (depending on different parameters) in parallel.

Here we make a first attempt toward such multi-diagnostic studies.  We
extend our 2D wind models from \one~to 3D, and relax the microclumping
approximation also for the optical mass loss
diagnostics \ha~and \heii\,$\rm \AA$ (Sect.~\ref{models}). In
Sect.~\ref{theor} we theoretically investigate \ha~and resonance line
formation in clumpy winds, and propose an analytic treatment of the
lines that does not rely on the microclumping approximation. A
simultaneous optical and UV diagnostic analysis is carried out in
Sect.~\ref{lcep_case} for the Galactic O6 supergiant \lcep, using
time-dependent radiation-hydrodynamic (RH) models as well as
stochastic ones together with our new tools for the radiative transfer
in clumped winds. These results are discussed in
Sect.~\ref{discussion}, while two initial applications of our analytic
formulation are given in Sect.~\ref{future}. We summarize the paper
and outline future work in Sect.~\ref{summary}.

\section{Wind models and radiative transfer}
\label{models}

\begin{table}
	\centering
	\caption{Parameters for the time-dependent 
                            RH model of \lcep~(see text).}
		\begin{tabular}{p{3.3cm}ll}
		\hline \hline Name & Parameter &
                Value \\ \hline Spectral type &  & O6 I(n)\,fp  \\ 
                Effective temperature & \teff & 36\,000\,K 
                \\  Stellar radius & \rstar & 21.1\,\rstun
                \\  Surface gravity & \logg & 3.55
                \\ Stellar luminosity & $\log L/L_\odot$ & 5.83  
                \\ Terminal speed  & \vinf & 2200\,\kms 
                \\ Mass-loss rate  & \mdot & $1.5\,\times$~\mdotun  
                \\ Helium abundance & $Y_{\rm He} \equiv n_{\rm He}/n_{\rm H}$ & 0.1 
                \\ CAK exponent  & $\alpha_{\rm CAK}$ & 0.7 
                \\ Initial Langevin  & $\varv_{\rm turb}/\varv_{\rm sound}$ & 0.5
                \\ turbulence fluctuation & &  
                \\
                \hline
		\end{tabular}
	\label{Tab:rh_par}
\end{table}

We create 2D and 3D RH and stochastic wind models by assembling
snapshots in radially independent wind slices (see Sect.~\ref{geom}).

Table 1 summarizes properties of a time-dependent RH model computed
following the approach in \citet{Feldmeier97}, which introduces base
perturbations from Langevin turbulence into an unstable line-driven
wind. The line force is computed with the nonlocal `Smooth Source
Function' \citep[SSF;][]{Owocki96} method that allows one to follow
the nonlinear evolution of the strong, intrinsic line-deshadowing
instability, while also accounting for the diffuse
line drag \citep{Lucy84} that reduces (and even eliminates) the
instability near the wind base. The net result is a highly structured
wind characterized by high-speed rarefactions and slower, dense clumps
(actually shells in these 1-D simulations). In comparison to
self-excited instability simulations \citep[e.g.,][]{Runacres02}, the
base perturbations here induce a somewhat lower onset and greater
velocity dispersion of the wind clumping. A central goal here is to
examine the effects of this extensive structure on wind
diagnostics. Stellar and wind parameters are taken from 
\citet{Repolust04}, except for the mass-loss rate (see Sect. 4).

Basic assumptions of our empirical, stochastic models were described
in detail in Paper I. Essentially, they are constructed so to resemble
the main structures predicted by the RH simulations, while still
allowing for a variation in the key parameters controlling the line
formation (see below).

\subsection{Parameters describing a structured wind}
\label{stoch_par}

\begin{table}
	\centering
	\caption{Basic structure parameters defining a stochastic wind model.}
		\begin{tabular}{lp{0.5cm}}
		\hline \hline Name & Parameter 
                \\ \hline Clumping factor$^a$ & $f_{\rm cl}$ 
                \\ Average time interval & $ \delta t$
                \\ between release of clumps &  
                \\ Interclump medium density parameter& $x_{\rm ic}$ 
                \\ Velocity span of clump  & $\delta \varv$  
                \\  
                \hline
                \\
                \multicolumn{2}{l}{$^a$~\fcl~may be replaced by the volume filling factor \fv.} 
                \\ \multicolumn{2}{l}{The two are related via \xic~(see \one).}
		\end{tabular}
	\label{Tab:stoch_par}
\end{table}

When creating our \textit{stochastic} wind models, we take an
heuristic approach and use a set of parameters to define the
structured medium. The clumping
factor \fcl($\varv$)\,$\equiv \langle \rho^2 \rangle/ \langle \rho \rangle^2$,
with the angle brackets denoting spatially averaged quantities, is the
only necessary structure parameter when calculating spectra via the
microclumping technique. Microclumping gives rise to the well known
result that the opacities for processes that depend on the square of
the density (for example \ha~emission in OB-stars) are augmented
by \fcl~as compared to a smooth model with the same mass-loss rate; in
contrast, opacities for processes that depend linearly on the density
(for example the {\sc uv} resonance lines) are not directly affected.
Thus, if the wind is clumped, mass-loss rates derived from smooth
models applied to \ha~are overestimated by a factor of $\sqrt{f_{\rm
cl}}$. In addition, the occupation numbers are modified for all
diagnostics because of the changed rates in the statistical
equilibrium equations. For a comprehensive discussion on the effects
of microclumping on various diagnostics, see \citet{Puls08}.

If the assumptions behind the microclumping model are not satisfied
(e.g. if clumps are optically thick for the investigated diagnostic),
the line formation will depend on more structure parameters than
just \fcl. Thus, relaxing the microclumping approximation means that
we must consider additional parameters when describing the structured
wind. These parameters (for a two component medium) were defined and
discussed in \one, and are listed in Table~\ref{Tab:stoch_par}. We
stress that they are essential for the radiative transfer in an
inhomogeneous medium, and not merely `ad-hoc parameters' used in a
fitting procedure.

In addition to the clumping factor \fcl~(or alternatively \fv),
\xic\,$\equiv \rho_{\rm ic}/\rho_{\rm cl}$ denotes the density ratio of the
interclump (ic) to clumped (cl) medium. The time interval \dt~(given
in units of the wind's dynamic time scale, \rstar/\vinf, and
not necessarily constant throughout the wind) effectively sets the
physical distances between clumps, also known as the porosity length
$h$ \citep{Owocki04}, which in our geometry is given by $h
= \varv_\beta \delta t$. Moreover, assuming a smooth underlying field
of customary $\beta$-type, $\varv_\beta(r) =
\varv_\infty(1 - b/r)^\beta$ with $b$ set by the assumed velocity at the
wind base $\varv_{\rm min} = \varv(r = 1)$, this time interval sets the {\it
velocity separation} between the clumps $\Delta \varv \approx
\varv_\beta \delta t d \varv_\beta / dr$ (Appendix A). Finally, the ratio
of the clump velocity span \dv~(as defined in Fig. 2 in Paper\,I) to
this velocity separation (representing a {\it velocity} filling
factor, see Appendix A) largely controls how strongly a perturbed
velocity field affects line formation\footnote{In this paper, we do
not consider the `jump velocity parameter', $\varv_{\rm j}$, defined
in Paper\,I, since it was shown there that this parameter mainly
influences the formation of very strong saturated lines, which are not
considered here. In our applications in Sect.~\ref{lcep_case}, we
simply set $\varv_{\rm j}/\varv_\beta=0.15$, which was found to be a
prototypical value in Paper\,I.}. In addition to these basic
parameters, the velocity $\varv_{\rm cl}$ (or radius $r_{\rm cl}$) at
which clumping is assumed to start also plays an important role for
the line formation. Note also that the parameters defining these
stochastic winds are independent of the physical origin to the
inhomogeneities.

The stochastic models should be distinguished from the time-dependent
RH simulations. In the latter the structure arises naturally from
following the time evolution of the wind and stems directly from the
line-driven instability. Thus, the time-averaged structure parameters,
as functions of radius, are an \textit{outcome} of these simulations
(in contrast to the stochastic models, where they are used as
fundamental parameters defining the structured wind). Nonetheless, the
exact wind structure still depends on the chosen initial conditions,
for example on whether the instability is self-excited or triggered by
some excitation mechanism (the latter is done here, see
Table~\ref{Tab:rh_par}). Finally, as shown in Paper\,I, by choosing a
suitable set of structure parameters one can reconcile spectrum
synthesis results stemming from the stochastic models with those from
RH simulations.

\subsection{Radiative transfer}

For resonance lines we use the Monte-Carlo code described in \one, but
a new radiative transfer code has been developed for the synthesis of
wind recombination lines presented here.  We investigate the O star
recombination lines \ha~and \heii. Recall that recombination lines and
resonance lines are formed differently. First, the optical depths are
calculated in different ways. For resonance lines, the optical depths
may be computed via a line-strength parameter, $\kappa_0$, which is
assumed to be constant throughout the wind and is proportional to the
product of the mass-loss rate and the abundance (by number),
$\alpha_{\rm i} \equiv n_{\rm i}/n_{\rm H}$, of the considered element
$i$. $\kappa_0$ may be expressed as \citep[e.g.,][]{Puls08}
\begin{equation} 
  \kappa_0 = \frac{\dot{M}}{R_\star \varv_\infty^2} \frac{\pi e^2/m_{\rm e}c}{4 \pi m_{\rm H}}
  \frac{\alpha_{\rm i}}{1+4Y_{\rm He}} f_{\rm lu} \lambda_0,  
  \label{Eq:kappa0_tot}
\end{equation} 
where $e$ and $m_{\rm e}$ are the electron charge and mass,
respectively, $c$ the speed of light, $m_{\rm H}$ the atomic hydrogen
mass, $Y_{\rm He}$ the helium abundance, $f_{\rm lu}$ the transition's
oscillator strength, and $\lambda_0$ its rest wavelength. The
advantage with this definition is that the radial Sobolev optical
depth in a smooth wind collapses to
 \begin{equation}
  \tau_{\rm Sob} = q \frac{\kappa_{\rm 0}}{r^2 \varv {\rm d} \varv/{\rm d}r},
  \label{Eq:tausob_app}
\end{equation}
where $q$ is the ionization fraction of the considered ion, and $r$ is
measured in units of \rstar~and $\varv$ in units of \vinf.

For \ha, the analog to $\kappa_0$ is the parameter
$A$ \citep[][Eqs.~1-3]{Puls96}.  $A$ is proportional to the mass-loss
rate \textit{squared} and to the NLTE departure coefficient, $b_{\rm
i}$, of the lower transition level (minus the correction factor for
stimulated emission). $b_{\rm i}=n_{\rm i}/n_{\rm i}^{\rm *}$, where
$n_{\rm i}^*$ is the occupation number of level $i$ in LTE with
respect to the ground state of the next ionization state.  In this
case, the radial Sobolev optical depth in a smooth wind
becomes \citep{Puls96}
 \begin{equation}
  \tau_{\rm Sob} = \frac{A(r)}{r^4 \varv^2 d \varv/dr}.
  \label{Eq:tausob_ha}
\end{equation}
In addition to their different optical depths, recombination lines are
(mainly) formed by recombining ions creating wind photons, whereas
resonance lines are formed by re-distributing photospheric stellar
continuum radiation by line scattering. Therefore the participating
atomic levels for recombination lines are rather close to LTE with
respect to the next ionization state (see Fig.~\ref{Fig:fcl}), which
means that the departure coefficients and thereby also the line source
function, $S_{\rm l} \propto (e^{h \nu /kT}b_{\rm l}/b_{\rm
u}-1)^{-1}$, for these lines are basically unaffected by the radiation
field and its dilution. In turn this allows us to prescribe the source
functions \citep{Puls96, Puls06} and simply carry out the `formal
integrals' within our stochastic and RH winds. In the present work,
we \textit{assume} that changes in the NLTE departure coefficients due
to optically thick clumps can be neglected for recombination-based
line formation, and simply calculate the $b_{\rm i}$'s from NLTE model
atmospheres using the microclumping approximation. Taking the example
of \ha~in O stars, this assumption should be reasonable, for
the \ha~departure coefficients in this domain are very close to unity
and the ionization of hydrogen is complete. However, for the case of,
e.g., A-supergiants, the assumption no longer holds, because in that
stellar domain \ha's lower level becomes the effective ground state of
hydrogen, which means that the line transforms to a quasi-resonance
line (and thereby that $S_{\rm l}$ depends on the radiation
field, \citealt{Puls98}).

As described in Paper\,I, also the ionization fractions $q$ for
resonance line formation are calculated assuming microclumping.  These
fractions are used both within clumps and for the interclump
medium. The potential feedback effects of optically thick clumping on
the departure coefficients and ionization fractions will be
investigated by incorporating the analytic methods developed in
Sect.~\ref{theor} into suitable NLTE atmosphere codes, and reported in
a future paper.

The assumption of prescribed departure coefficients is an enormous
simplification compared to the UV resonance lines, and has enabled us
to extend our 2D wind models to 3D when modeling recombination
lines. In the synthesis we follow the basic method introduced
by \citet{Puls96}, which does not rely on the Sobolev approximation,
with appropriate modifications for the line opacities of \heii. A
core/halo approach is adopted, in which a photospheric profile is used
as a lower boundary input (at $r=1$, with $r$ in units of the stellar
radius) and the radiative transfer is solved only in the wind. As for
the resonance lines (Paper\,I), we assume pure Doppler line broadening
within the wind, characterized by a thermal speed, $\varv_{\rm t}$,
given throughout the paper in units of \vinf.

Our new recombination line code has been extensively tested and showed
to yield equivalent results with \citet{Puls06} for smooth
winds. Also, results based on the microclumping technique are
reproduced for stochastic as well as RH winds with low wind densities,
as expected because the clumps then remain optically thin. In our
applications, we use hydrogen and helium occupation numbers calculated
by {\sc fastwind} model atmospheres \citep{Puls05}, under the
microclumping approximation, as input for the radiative transfer to
compute synthetic spectra. Photospheric profiles are taken from NLTE
calculations of atmospheres with negligible winds.  The consistency
between unified (meaning a simultaneous treatment of the photosphere
and wind) model atmosphere calculations and the simplified core/halo
approach has been verified in the microclumping limit, for
recombination lines as well as for resonance lines. Moreover, we have
found that averaged recombination line profiles calculated from our
earlier 2D, stochastic models are almost identical to those calculated
from our new 3D ones, as was already anticipated for the UV resonance
lines in \one.

\paragraph{The He\,{\sc ii} blend in \ha.} 
The star's helium abundance has of course been considered in the
calculation of the \ha~wind opacity, but for simplicity we include the
He\,{\sc ii} blend only in the photospheric profile, thus neglecting
its direct contribution to the wind emission.  This results in a
slight underestimate of the total wind opacity of the line
complex. However, by comparing to unified model atmosphere
calculations that consistently treat the He\,{\sc ii} blend, we have
found that the direct helium contribution is low for our typical stars
of interest, and in our applications for \lcep~it can even be
neglected.  Although sufficient for our purposes here, this approach
should obviously not be generalized, because it may yield unrealistic
results for stars with parameters different from our template star.
     
\subsection{Geometry}
\label{geom}

\begin{figure}
  \resizebox{\hsize}{!}{\includegraphics[angle=90]{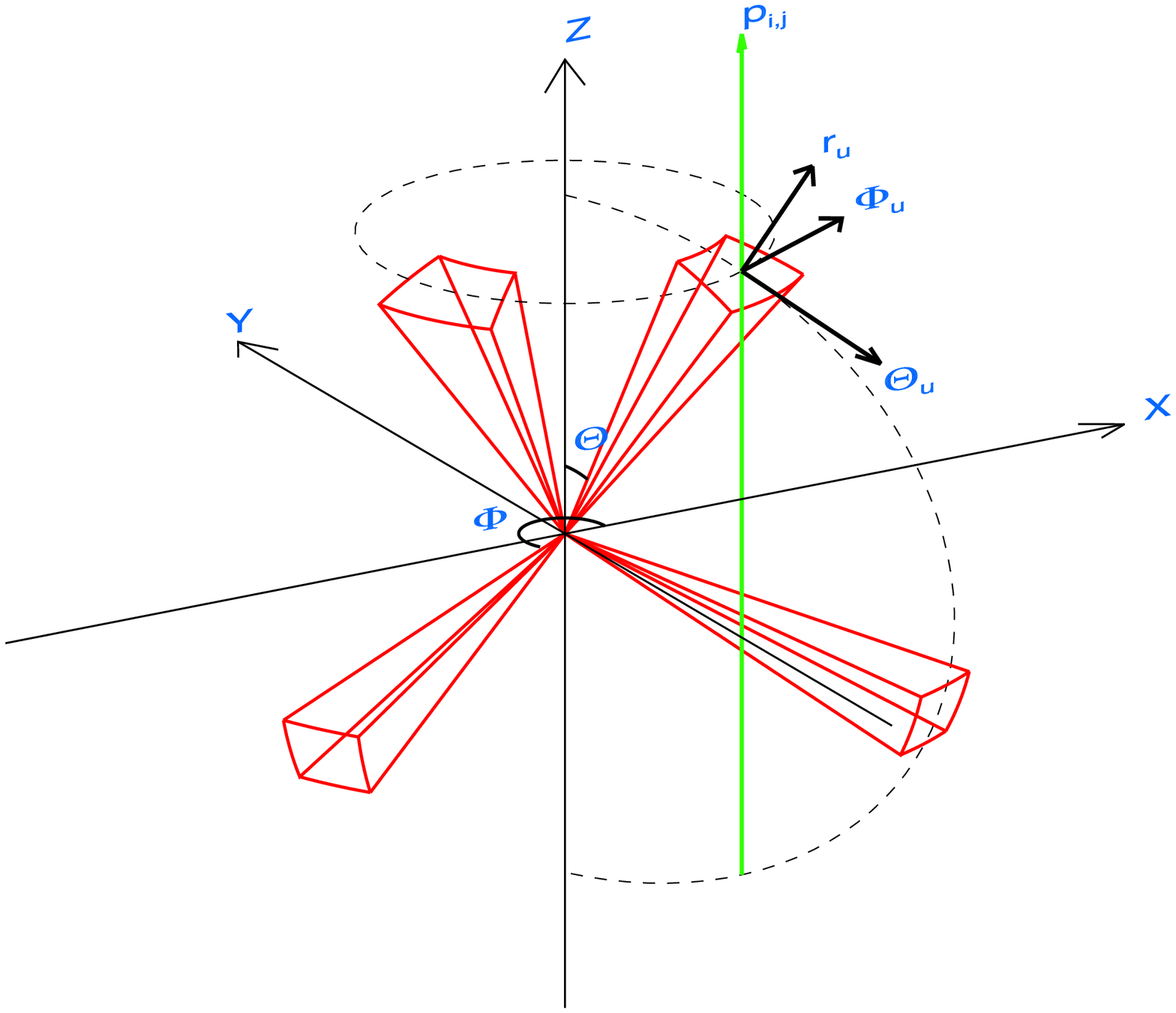}}   
   \caption{Illustration of the wind geometry, see text. \textit{A
color version of this figure is available in the web version.}}
\label{Fig:geometry}
\end{figure}

To construct (pseudo-)3D winds, we use the `patch method' from
\citet{Dessart02}. A standard right-handed spherical system
($r,\Theta,\Phi$) is used, defined relative to a Cartesian set
($X,Y,Z$). However, when computing recombination lines, we no longer
assume symmetry in the azimuthal ($\Phi$) direction (as was done
in \one). The lateral scale of coherence in the wind is set by the
parameter $N_\Theta$ and by assuming that the physical coherence
lengths in both lateral directions are approximately equal
(Fig.~\ref{Fig:geometry}). This assumption is reasonable because,
within our approach, which for example does not include an axis of
rotation, all observer directions should be alike. Thus, if we desire
a coherence scale of 3 degrees, the number of slices in the polar
direction should be $N_\Theta=180/3=60$ and in the azimuthal direction
$N_\Phi = \rm int$$[2 N_\Theta \sin \Theta]$, i.e., $2 N_\Theta$ at
the equator but fewer toward the pole in order to preserve
the \textit{physical} length scales. Wind slices are then assigned
randomly from a large number of spherically symmetric simulations
(either RH or stochastic).

$N_\Theta$ thus enters all our models as an input parameter.  Paper\,I
showed that this parameter does not change the strengths of the
resonance lines. More tests have shown that also the effects on
recombination lines are modest for investigated values. Therefore all
3D models in this paper assume $N_\Theta=60$, meaning a coherence
length of 3 degrees at the equator, which is consistent with
observational constraints derived from line-profile variability
analysis \citep{Dessart02}. Theoretical constraints on $N_\Theta$ are
still lacking, and will require a careful treatment of the lateral
radiation transport in RH models. The first 2D simulations
by \citet{Dessart03} neglected this transport and resulted in a
laterally fragmented wind down to the grid scale but the follow-up
study \citep{Dessart05} included a simplified 3-ray approach and
resulted in larger (but un-quantified) lateral coherence scales.

For recombination line formation, the observer is assumed to be
located at infinity in the $Z_{\rm u}$ (subscript u denoting a unit
vector) direction.  The geometry is sketched in
Fig. \ref{Fig:geometry}. We solve the radiative transfer using a
traditional $(P,Z)$ system for a set of P-rays, each defined by the
minimum radial distance to the $Z$ axis and by the azimuthal angle
$\Phi$, which is constant along a given ray. If the angle between the
ray and the radial coordinate is $\theta$, then $\mu = \cos \theta$
and $P = r \sqrt{1-\mu^2}$. Thus, for rays in direction $Z_{\rm u}$
the radiation angle $\theta$ coincides with the polar coordinate
$\Theta$, and it becomes trivial to calculate the physical locations
at which wind-slice borders are crossed. The observed flux may then
finally be computed by performing a double integral of the emergent
intensity over $P$ and $\Phi$.

\section{Theoretical considerations of resonance and recombination 
line formation in clumpy winds}
\label{theor}

Resonance line formation in clumpy hot star winds was discussed in
detail in \one. There we identified an intrinsic coupling between the
effects of porosity and vorosity \citep[='velocity
porosity',][]{Owocki08}, which we here further elaborate upon. In
particular, we propose an analytic formulation of line formation in
clumped hot star winds (that does not rely on the microclumping
approximation). As already mentioned in Sect.~\ref{models}, the
development of such simplified approaches is important for properly
including effects of optically thick clumping into atmospheric NLTE
codes. For recombination lines, we focus on \ha~and discuss impacts
from optically thick clumping on its formation, using our stochastic
wind models as well as an extension of the analytic treatment
developed for the resonance lines.

\subsection{Analytic treatment of resonance lines in clumpy winds}
\label{analytic}

\begin{figure}
  \begin{minipage}{8.0cm} \resizebox{\hsize}{!}
        {\includegraphics[angle=90]{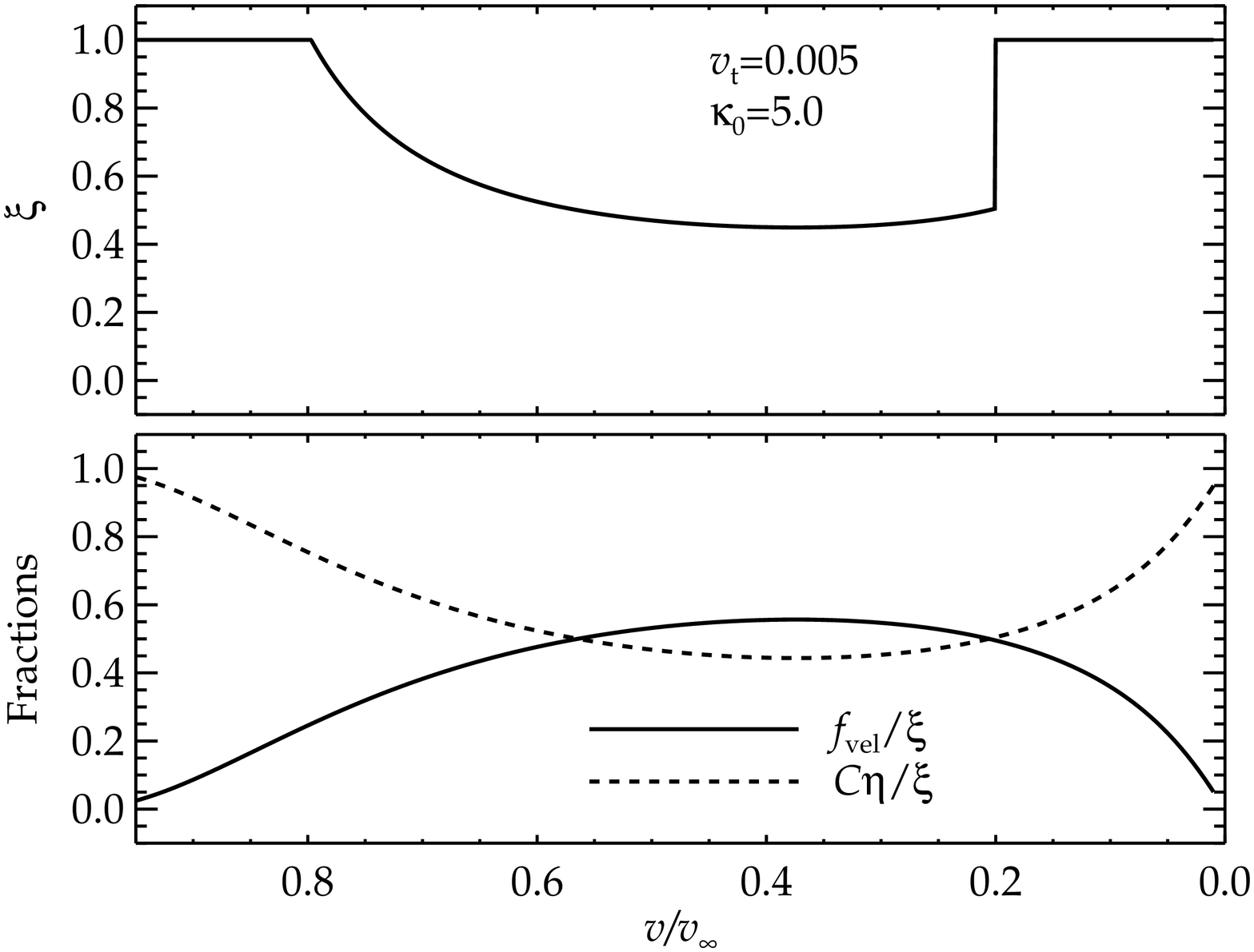}} 
    \end{minipage} 
    \begin{minipage}{8.0cm} 
        \resizebox{\hsize}{!}
        {\includegraphics[angle=90]{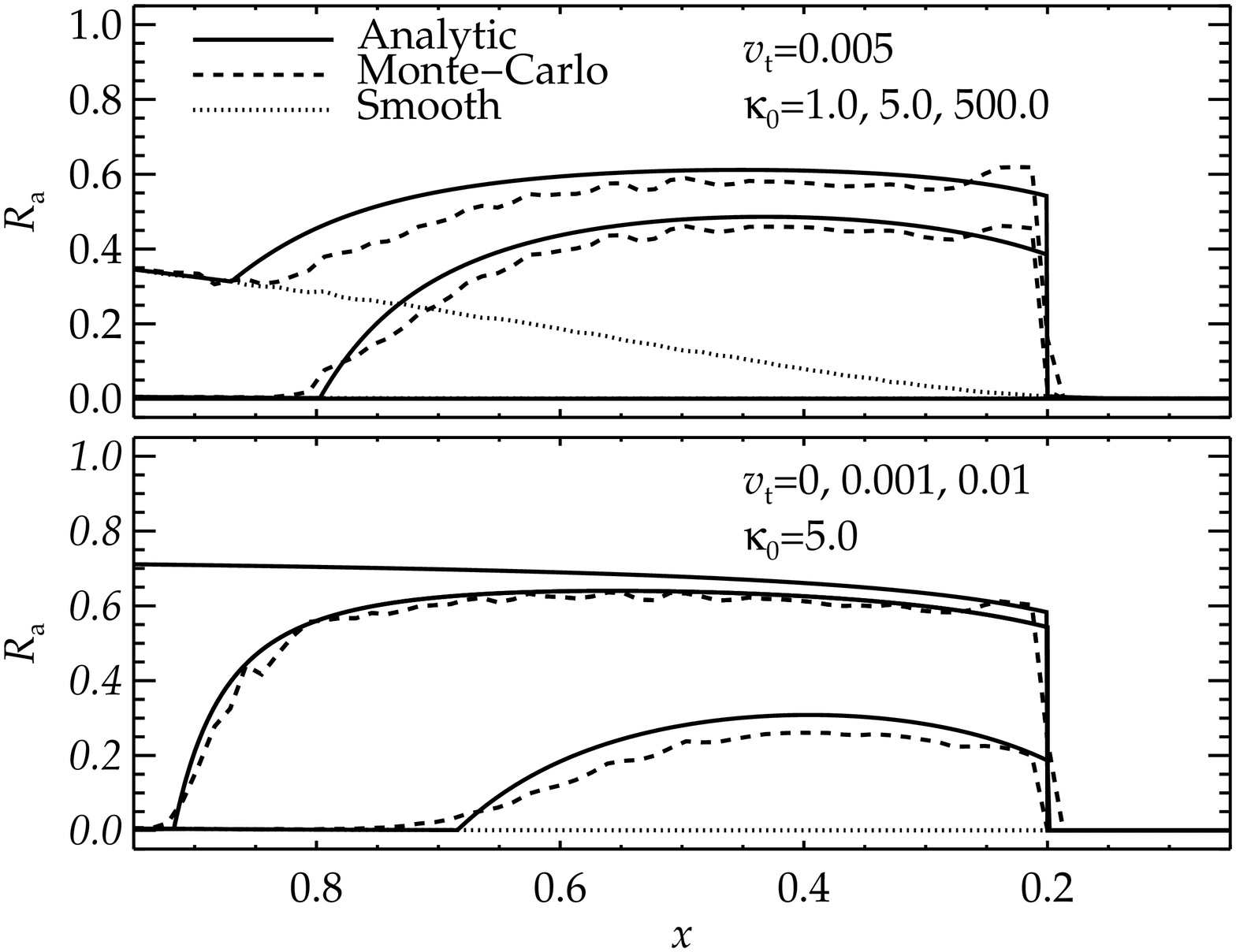}} 
    \end{minipage} 
   
\caption{
  \textit{Panel a:} $\xi$ (Eq.~\ref{Eq:xi}) as a function of wind
  velocity. \textit{Panel b:} The relative contributions to $\xi$
  from \fvel~and $C \eta$. Curves in panels $a$ and $b$ have been
  calculated with $\kappa_0=5$ and thermal speed $\varv_{\rm t}=0.005$
  (in units of \vinf). \textit{Panels c and d:} Analytic
  (Eq.~\ref{Eq:Fow_1}, solid lines) and Monte-Carlo (dashed lines)
  based absorption part resonance line profiles from clumped winds, as
  compared to smooth results (dotted lines). Clumping starts at
  $\varv_{\rm cl}=0.2$ ($r_{\rm cl}=1.24$). \textit{Panel
  c:} Profiles for three different values of the 
  line strength $\kappa_0$ (indicated in the
  figure), with increased absorption for higher values of
  $\kappa_0$. Only the $\kappa_0=1$ profile is not
  saturated (i.e., does not display $R_{\rm a}=0$) for smooth
  models. \textit{Panel d:} Profiles for $\kappa_0=5$ and different
  values of $\varv_{\rm t}$, as indicated in the figure. No
  Monte-Carlo profile for the $\varv_{\rm t}=0$ case is shown (simply
  because we have not yet developed a `Sobolev version' of this
  code). Analytic and Monte-Carlo profiles display increased absorption
  for higher values of $\varv_{\rm t}$, whereas all smooth profiles
  have $R_{\rm a}=0$.}
\label{Fig:analytic}
\end{figure}

Throughout this section we assume a \textit{smooth velocity
field}, characterized by $\beta=1$ (Sect.~\ref{stoch_par}).
Despite this, the vorosity effect will be demonstrated to be important
for the line formation (i.e., a non-monotonic velocity field is not
required for vorosity to be at work). Adopting the normalized,
dimensionless frequency $x=(c/\varv_\infty)(\lambda_0 - \lambda)
/ \lambda$, and following the basic arguments of \citet{Owocki08}, we
write the absorption part of a normalized resonance line profile,
$R_{\rm a,x}$, from a radial ray as (see Appendix A)

\begin{equation}
  R_{\rm a,x} = \xi_{\rm x} e^{-\tau_{\rm cl,x}} + (1-\xi_{\rm x}) e^{-\tau_{\rm ic,x}}.
  \label{Eq:Fow_1}
\end{equation} 

\noindent $R_{\rm a,x}$ describes the part of the profile that stems from
absorption of continuum photons released from the photosphere. The
\textit{total} line profile is given by $R_{\rm x} = R_{\rm a,x} + R_{\rm em,x}$, 
where $R_{\rm em,x}$ is the re-emission profile. The scattering nature
of the resonance line source function significantly complicates the
formation of $R_{\rm em,x}$. Therefore, for now we restrict ourselves
to discussing $R_{\rm a,x}$ from a radial ray for these lines. For
recombination lines, on the other hand, we will lift these
restrictions and treat the complete line profile (Sect.~\ref{ha}).
Obviously this must be done also for resonance lines before, e.g.,
including our analytic formalism in NLTE model atmosphere codes (see
also Sect.~\ref{doublets}). In any case, recall that $R_{\rm a,x}$
controls the actual line-profile strengths of resonance lines, because
these are pure scattering lines formed out of re-distributed continuum
radiation emerging from the photosphere.

In Eq.~\ref{Eq:Fow_1} we define $\xi$ as the \textit{fraction of the
velocity field over which photons may be absorbed by clumps}, with the
$\tau$'s representing the optical depths for the clumped (subscript
cl) and rarefied (subscript ic) regions. $\xi$ describes the essential
effects of Owocki's vorosity; the first term in Eq.~\ref{Eq:Fow_1}
handles the part of the line profile emerging from absorption within
the clumps, whereas the second term handles the part emerging from
absorption within the interclump medium. What remains then is finding
an appropriate expression for $\xi$. In Appendix A we argue that a
reasonable approximation may be

\begin{equation}
  \xi_{\rm x} \approx \frac{\delta \varv}{\Delta \varv} + C \frac{\varv_{\rm t}}{\Delta \varv} = f_{\rm vel}(\varv) + C(\varv) \eta(\varv),
    \label{Eq:xi} 
\end{equation} 

\noindent with \Dv~the velocity gap between two clump centers, \fvel~the
\textit{velocity filling factor} (defined in full analogy with the traditional 
volume filling factor), $\eta$ the \textit{effective escape ratio}
(here re-defined from \one, see Appendix A), and $C$ a correction
factor that depends on the line strength. The condition for
interaction for a radial line photon is in the Sobolev approximation
simply $x=\varv$, and, for simplicity, we from now on suppress all
velocity dependencies of the quantities in Eq.~\ref{Eq:xi}.  As shown
in Appendix A, we may write

\begin{equation}
  f_{\rm vel} = f_{\rm v} |\frac{\delta \varv}{\delta \varv_\beta}|, \qquad 
  \eta = \frac{L}{h},
  \label{Eq:3}
\end{equation}

\noindent where $h$ is the porosity length of the medium
and $L=\varv_{\rm t}/(d \varv/dr)$ the (in this case radial) Sobolev
length. For the smooth velocity field considered in this section,
$\delta \varv = \delta \varv_\beta$, which gives \fvel=\fv. Even
though the principle effect of the optically thick clumps on resonance
line formation is a velocity effect governed by \fvel,
Eqs.~\ref{Eq:Fow_1}-\ref{Eq:3} indicate there is also a dependence on
spatial porosity through the ratio $\eta = L/h$.  This coupling was
argued for already in \one. However, it appears that $\xi$ better
characterizes the effects of clumping in resonance line formation than
did our previous parametrization (see Appendix A).  The optical depths
entering Eq.~\ref{Eq:Fow_1} may be approximated by the corresponding
Sobolev ones, corrected for the influence of $\eta$ (Appendix~A). Then
for resonance lines, $\tau_{\rm cl,x} \approx \tau_{\rm sm,x}/\xi_{\rm
x}$ and $\tau_{\rm ic,x} = \tau_{\rm cl,x}x_{\rm ic}$, where
$\tau_{\rm sm,x}$ is given by Eq.~\ref{Eq:tausob_app}.  Note that all
parameters used to define our stochastic wind models
(Table \ref{Tab:stoch_par}) enter the expression for $R_{\rm a,x}$,
illustrating that indeed all these are important for the general line
formation problem.

The upper two panels of Fig.~\ref{Fig:analytic} plot $\xi$ as well as
the relative contributions from \fvel~and $C \eta$ for a resonance
line with line-strength parameter $\kappa_0=5$. We do not allow for
$\xi$ values above unity (important for the outer wind, see also
Appendix A), and we also recover the smooth result for velocities
lower than $\varv_{\rm cl}$ simply by setting $\xi_{\rm
x \le \varv_{\rm cl}}=1$. All models displayed in
Fig.~\ref{Fig:analytic} were calculated with density structure
parameters \fcl=4.0, \dt=0.5, and \xic=0.0025. For a \textit{smooth}
model (with ionization fraction $q=1$, assumed in this section),
$\kappa_0=5$ results in a profile at the saturation threshold.  In the
lower two panels we show analytic absorption-part line profiles
calculated using Eq.~\ref{Eq:Fow_1} and profiles calculated using our
Monte-Carlo code.  To make consistent comparisons between methods, we
accounted only for radial photons in the Monte-Carlo simulations.  The
agreement between the methods is very good, lending support to the
proposed analytic treatment and providing a relatively simple
explanation for the basic features of the synthetic profiles.

Evidently, profile-strength reductions can be quite dramatic for
`moderately strong' cases such as $\kappa_0=5$. For the very strong
$\kappa_0=500$ line also the interclump medium is optically thick and
the profiles are therefore saturated (which is a necessity because
such saturated profiles are observed in hot stars). Note that,
if \dv~were much higher than $C \varv_{\rm t}$, one could neglect the
second term in Eq.~\ref{Eq:xi} and $\xi$ would become independent of
the porosity length. If one also neglects the interclump medium
(setting \xic=0), and assumes that clumps are optically thick
throughout the entire wind (appropriate for the $\kappa_0=5$ line),
then the observer in our example would simply receive a constant
residual flux $R_{\rm a,x}=1-f_{\rm v}=0.75$.
Fig.~\ref{Fig:analytic}d shows that this generally does not hold (even
for the idealized case of zero thermal speed, the interclump medium
still plays a role), demonstrating that, along with the velocity
filling factor $f_{\rm vel}$, in general both \xic~and $\eta$ also
help shape the emergent profile for a wide range of line strengths and
structure parameters. Fig.~\ref{Fig:analytic}d illustrates the
importance of accounting for the finite line profile width. $C \eta$
may not be neglected, even in models with very low, but finite thermal
velocity, and becomes particularly important toward the blue edge of
the line profiles.  This occurs because the resonance zones in the
outermost wind become very radially extended. $L$ thus grows whereas
the distances between the clumps (determining $h$) are unaffected due
to the very slowly changing velocity field. Consequently $\eta$
becomes very high and $\xi$ eventually reaches unity. Since the smooth
$\kappa_0=5.0$ line is optically thick, a `blue absorption dip'
(extensively discussed in \one) is created.

Randomization effects are here neglected because we have used a smooth
velocity field. When clumps are allowed to have velocities higher and
lower than those given by the mean velocity field, overlapping
velocity spans of the clumps lead to increased escape of photospheric
photons. The blue absorption dip then becomes less prominent than what
is displayed in Fig.~\ref{Fig:analytic}, as discussed in \one~(see
also Appendix A, for some comments on randomization effects).

Nevertheless, this section demonstrates that the microclumping
approximation can result in large errors if indeed the wind is clumped
but the clumps are not optically thin. First applications of the
analytic formulation are given in Sect.~\ref{future}, for diagnostics
of \textit{weak wind stars} and for the predicted profile-strength
ratios in resonance line \textit{doublets}.

\subsection{Recombination lines in clumpy winds}
\label{ha}

\begin{figure}
  \begin{minipage}{8.0cm} \resizebox{\hsize}{!}
        {\includegraphics[angle=90]{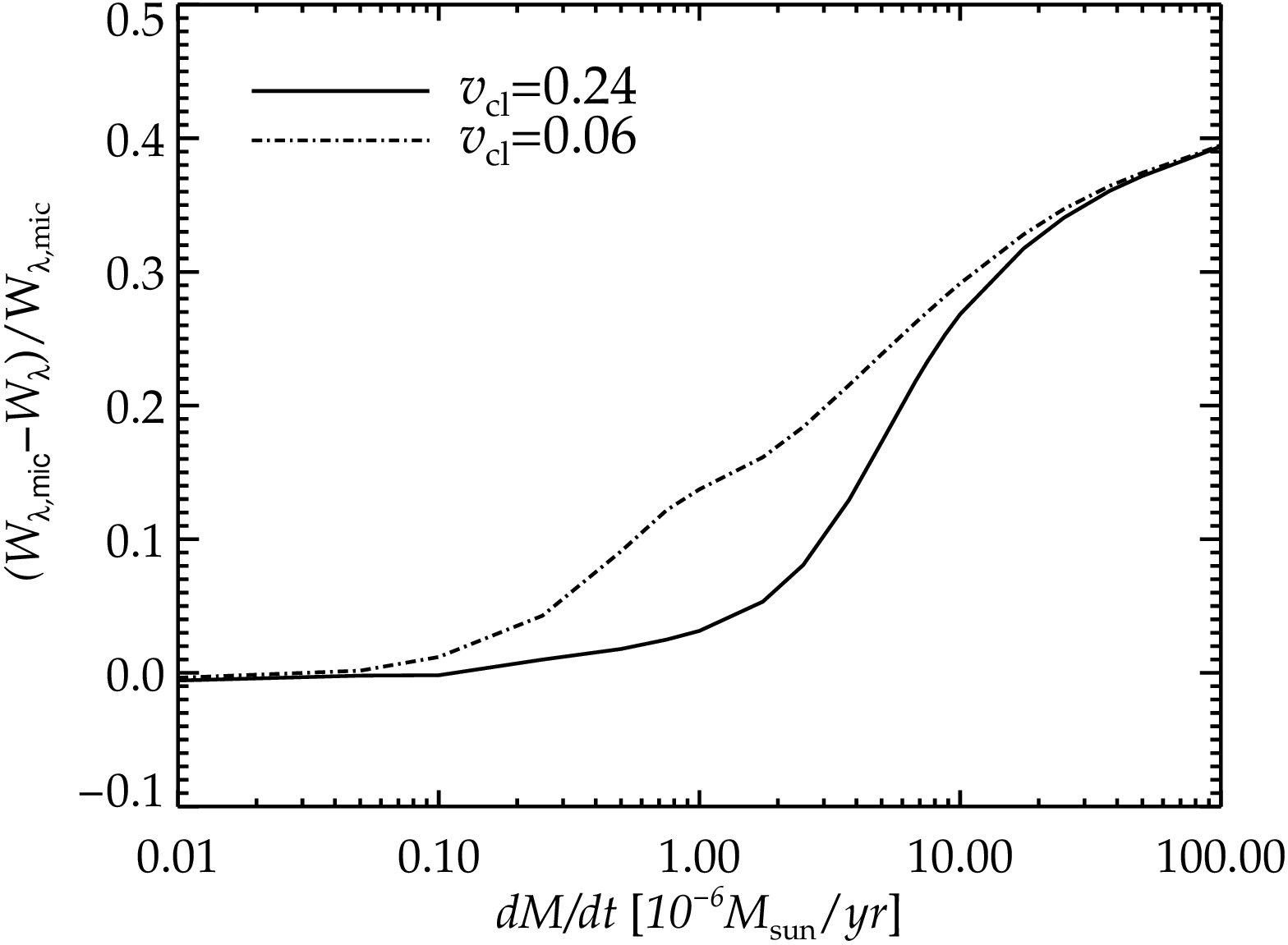}} 
    \end{minipage} 
    \begin{minipage}{8.0cm} 
        \resizebox{\hsize}{!}
        {\includegraphics[angle=90]{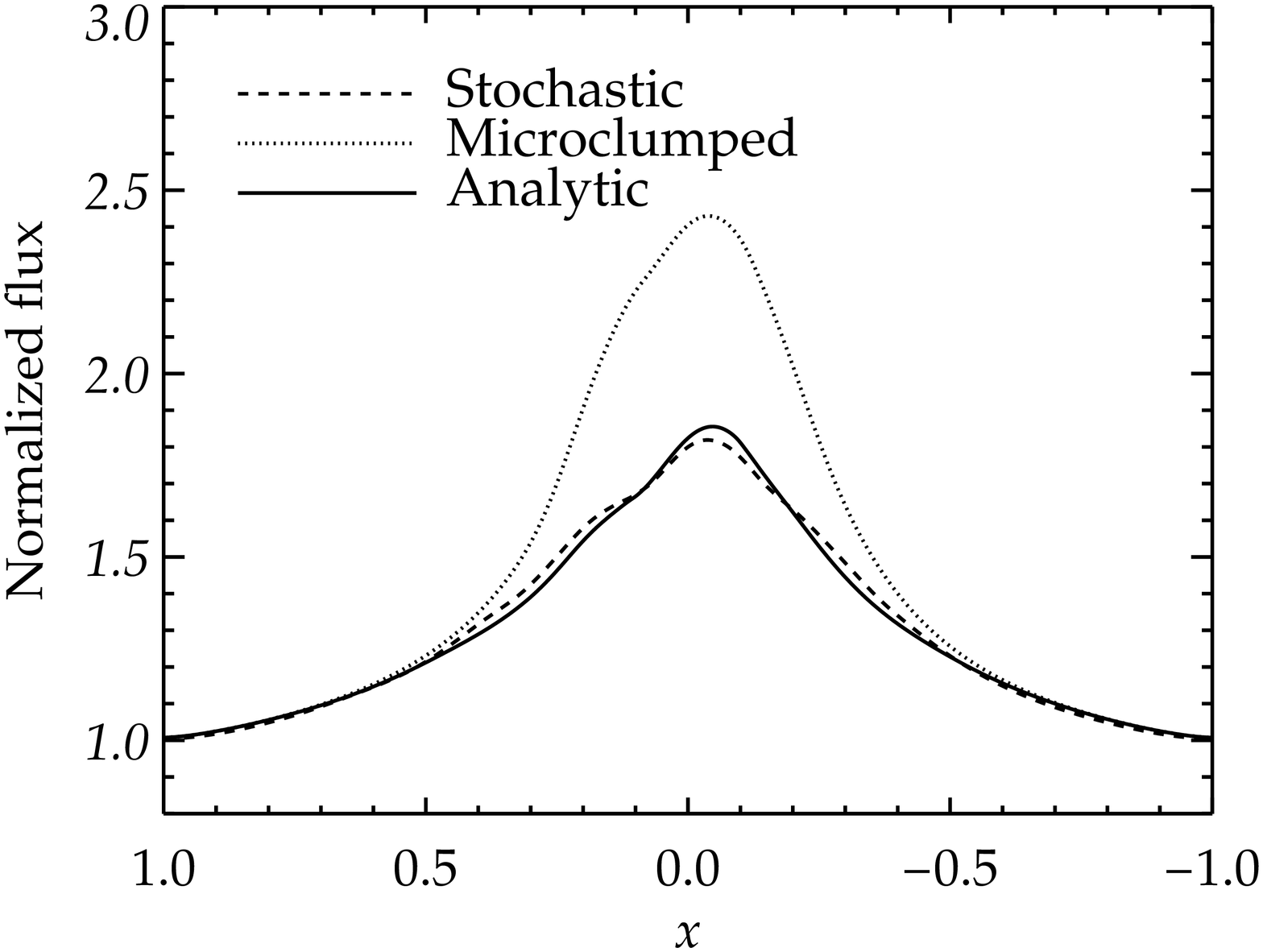}} 
    \end{minipage} 
\caption{
    \textit{Upper panel:} Deviations from the microclumping
    approximation of equivalent widths of synthetic \ha~ line profiles
    versus mass-loss rate (see text). Clumping starts at $\varv_{\rm
    cl} = 0.24$ and 0.06 ($r_{\rm cl}=1.3$ and 1.05), respectively,
    as indicated in the figure. \textit{Lower panel:} \ha~line
    profiles as calculated by stochastic, analytic, and microclumped
    models with \fcl=9 and \mdot=10\,$\times$\,\mdotun, and the rest of
    the stellar and wind parameters as for \lcep. Clumping for all
    models starts at $\varv_{\rm cl}=0.24$.}
\label{Fig:ha_ew}
\end{figure}

We now leave the resonance lines behind and turn to the formation of
recombination lines. We focus on \ha, the primary spectroscopic
mass-loss diagnostic for O stars. \heii~reacts similarly as \ha~to
clumping in our primary stars of interest (because He\,{\sc iii} is
the dominant ion in the line forming regions), and will be considered
only in our diagnostic study of \lcep~(Sect.~\ref{lcep_case}).

First we present results from calculating \ha~line profiles using our
stochastic 3D wind models. Our main interest is to investigate
differences with respect to the microclumping model, so main results
are provided in terms of the deviation of the equivalent
widths \wl~between the two methods, $(W_{\lambda,\rm
mic}-W_{\lambda})/W_{\lambda,\rm mic}$, as a function of mass-loss
rate (here $W_{\lambda,\rm mic}$ denotes \wl~as calculated from a
model assuming microclumping). All models discussed in this section
were calculated with unity departure coefficients, wind electron and
radiation temperatures as for approximately \lcep~\citep[calibrated
using unified NLTE model atmospheres, see][]{Puls06}, and no input
photospheric absorption profiles. We used structure
parameters \fcl=9.0, \dt=0.5, $x_{\rm ic}=0.0025$, and a smooth
velocity field characterized by $\beta=1$.

For typical O-supergiants, the equivalent widths of profiles
calculated from stochastic models are slightly lower than those based
on the microclumping technique
(Fig.~\ref{Fig:ha_ew}). Deviations stem from optically thick
clumps. The dominating effect is on the wind \textit{emission}
of \ha~photons rather than on the wind absorption of photospheric
photons (in contrast to resonance lines, see previous section).
This is because the source function for recombination lines is
basically unaffected by the dilution of the radiation field, which for
relatively strong and hot winds make these lines appear in emission
and thereby suffer the main effect from a clumped wind on the emission
part of the line profile. Moreover, the $\rho^2$-dependence of
recombination line opacity increases the contrast between the optical
depths for the clumps and those for the interclump medium, as
compared to resonance line formation. This lowers the significance of
the interclump medium and also causes the clump optical depths to
decrease faster for increasing radii. The latter effect results in
clumps that are optically thick only in the lower wind regions.
Deviations from the microclumping limit are therefore more significant
for cases with earlier onset of clumping.  For example, the equivalent
widths for the models with \mdot=2.5\,$\times$\,\mdotun~are reduced by
7\,\% and 17\,\% when clumping starts at $\varv_{\rm cl}=0.24$ and
$\varv_{\rm cl}=0.06$, respectively.  The effect is thus modest, but
noticeable. Remember that reductions are measured against models
assuming microclumping; the profiles are still much stronger than
profiles computed from smooth models with the same mass-loss rate.
  
Our tests show that effects are confined to the line core and that the
microclumping approximation provides accurate results in the line
wings. However, Fig.~\ref{Fig:ha_ew} reveals prominent emission
strength reductions for stronger winds, since then optical depth
effects become important for ever larger portions of the total wind
volume. Furthermore, the onset of clumping is irrelevant in these
strong winds because the majority of the emission emerges from radii
greater than $r_{\rm cl}$. This insensitivity to the onset of clumping
also recovers the scaling invariant for \textit{microclumped} winds
($\propto \sqrt{f_{\rm cl}} \dot{M}$, see Sect.~\ref{stoch_par}).  For
typical OB-supergiants, however, this scaling does not hold because of
the strong opacity contrast between wind radii lower than and greater
than $r_{\rm cl}$. Very strong recombination lines, such as those
displayed in Fig.~\ref{Fig:ha_ew} (lower panel), are typically seen in
WR stars, for which however also a reduced hydrogen content is
expected (as well as a breakdown of our assumption of an optically
thin continuum). Nonetheless, our analysis could, of course, be
generalized to recombination lines of other chemical species (as has
been done for \heii~in our application to $\lambda$ Cep), and may
point to significant optical depth effects in the strong emission
peaks of stars with very high mass-loss rates. Indeed, lower emission
peaks in the theoretical spectrum of a WR star have been found
by \citet{Oskinova07}, on the basis of scaling smooth opacities using
a porosity formalism.  However, when deriving empirical mass-loss
rates from microclumping models of WR stars one normally considers
also the electron scattering wings \citep[which are unaffected by
microclumping, see][]{Hillier91}, and because clumps in these probably
are optically thin it may be that lower emission peaks would have a
greater effect on the inferred clumping factors than on the mass-loss
rates.
  
\paragraph{Analytic treatment of recombination lines.} We can understand 
the reduction in \ha~emission strengths using the same analytic
treatment as outlined for resonance lines. Better yet, because the
source function $S$ is almost unaffected by the radiation field (see
Sect.~\ref{models}), we can for recombination lines simulate the total
profile, $R_{\rm x}=R_{\rm a,x}+R_{\rm em,x}$, writing

\begin{equation}
  R_{\rm em,x} = S \xi_{\rm x} (1 - e^{-\tau_{\rm cl,x}}) + S(1-\xi_{\rm x})(1 - e^{ -\tau_{\rm ic,x}}),  
  \label{Eq:reem_pap}
\end{equation} 

\noindent where $S$ is given in units of
the continuum intensity and evaluated at the resonance point. $R_{\rm
em,x}$ is much more influenced by non-radial photons than is $R_{\rm
a,x}$, so accordingly the radial streaming assumption from the
previous section must be relaxed here. Details are given in
Appendix A. Moreover, the clump and interclump optical depths now
become $\tau_{\rm cl,x} \approx \tau_{\rm sm,x}(f_{\rm cl}/\xi_{\rm
x})$ and $\tau_{\rm ic,x} = \tau_{\rm cl,x} x_{\rm ic}^2$, due to the
$\rho^2$-dependence of the line opacity, with $\tau_{\rm sm,x}$ given
by Eq.~\ref{Eq:tausob_ha}. Already in the previous paragraph we
mentioned how this lowers the significance of the interclump medium in
recombination line formation. Actually, tests have shown that, in our
typical stars of interest, the optical depths in the interclump medium
are so low that the second term in Eq.~\ref{Eq:reem_pap} can safely be
neglected. The lower panel of Fig.~\ref{Fig:ha_ew} illustrates that
profiles computed using the analytic approximation agree very well
with those computed using our stochastic wind models.

\section{A multi-diagnostic study of \lcep}
\label{lcep_case}

\begin{table}
  \centering
  \caption{Structure parameters for an empirical stochastic wind model of \lcep.}
  \begin{tabular}{lllll}
    \hline \hline
    Velocity range [$\varv_\beta$/\vinf] 
    & \fcl  
    & \dt
    & \xic 
    & \dv/\dvb  
    \\         
    \hline
    $\varv_{\rm min}$~-~0.15 & 1.0  &     & 1.0    &   ~1.0
    \\ 0.15~-~0.35         & 28.0 & 0.5 & 0.005  &  -5.0
    \\ 0.35~-~0.60         & 14.0 & 0.5 & 0.0025 &  -5.0
    \\ 0.60~-~0.95         & 14.0 & 3.0 & 0.0025 &  -5.0
    \\ 0.95~-~1.0          &  4.0 & 3.0 & 0.0025 &  -5.0
    \\
    \hline
  \end{tabular}
  \label{Tab:lcep_stoch}
\end{table}

We have carried out a detailed study of the Galactic O6
supergiant \lcep. This star was chosen in part to connect with 
\one~and in part because it is a well observed and studied object, 
with significant mass loss, that appears to be less peculiar than,
e.g., \zpup. A simultaneous investigation of optical diagnostics and
the \pv~resonance lines is performed. The ionization fractions of
\pv~and the hydrogen and helium departure coefficients (see
Fig.~\ref{Fig:fcl}) are calculated with the unified model atmosphere
code {\sc fastwind}, under the microclumping approximation and
assuming the same (smoothed) clumping factors as in corresponding RH
or stochastic models, with stellar and wind parameters as given in
Table~\ref{Tab:rh_par}, and with a solar \citep{Asplund05b} phosphorus
abundance. We account for stellar rotation by convolving the emergent
synthetic line profiles with a constant $\varv \sin i = 220\,\rm
km\,s^{-1}$, which is the photospheric value. Thus we neglect
differential rotation in the wind and simply assume that stellar
rotation only affects the line formation in the wind in such a way
that we may approximate the same $\varv \sin i$ value as for the
stellar photosphere (see also Sect.~\ref{m_est}).  Generally, for the
line profiles studied here, the influence of rotation is important for
the recombination lines, but not for the resonance lines. We use
observed {\sc uv fuse} spectra from \citet{Fullerton06}, and optical
spectra from \citet{Markova05} and A.~Herrero \citep[described
in][]{Herrero00}. In addition to $\rm H_\alpha$, \heii, and PV, we
also consider the wind sensitive cores of $\rm H_\beta$ and $\rm
H_\gamma$. However, for these diagnostics we rely entirely on the
microclumping approximation, which because of their low wind optical
depths should be sufficient.

\subsection{Clump optical depths}
\label{taucl}

\begin{figure}
  \resizebox{\hsize}{!}{\includegraphics[angle=90]{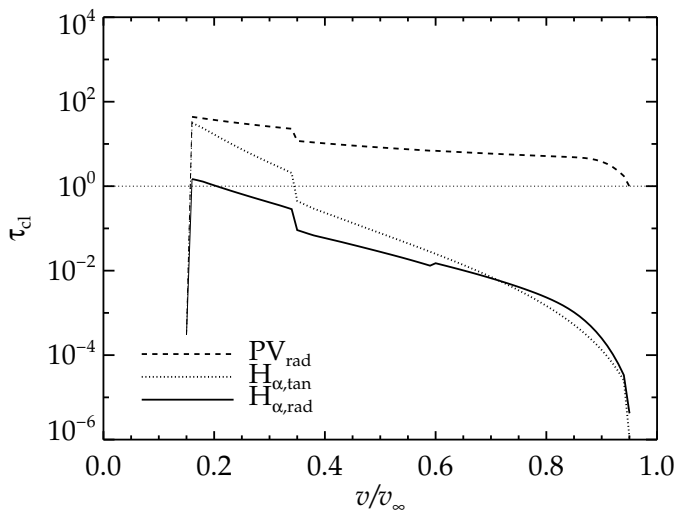}} 
  \caption{Radial clump optical depths for PV, and radial and
  tangential ones for \ha, as functions of wind velocity for our
  empirical, stochastic wind model of \lcep~(Table~\ref{Tab:lcep_stoch}).}
\label{Fig:taucl}
\end{figure}

The \textit{clump optical depth} in the wind is the primary quantity
governing the validity of the microclumping approximation. In Appendix
A (see also the previous section) we have provided estimates of the
clumps' radial \textit{Sobolev} optical depths in resonance and
recombination lines. However, in our stochastic models, clumps do not
always cover a complete resonance zone, so the Sobolev optical depths
must be replaced by optical depth calculations including the actual
line profile (cf. Paper\,I, Eq.~A.12, and note also that these `exact'
optical depth calculations then involve $f_{\rm vel}$ rather than
$\xi$, see Appendix A). Within our stochastic wind models, the radial
extent of a clump is $l_{\rm r} = \varv_\beta$\,\dt\,\fv, and
therefore, by transforming to the corresponding velocity width, we may
readily calculate the `actual' clump optical depth \taucl.

Fig.~\ref{Fig:taucl} shows \taucl~for \pv~and \ha, as calculated from
our empirical, stochastic model
of \lcep~(Table~\ref{Tab:lcep_stoch}). The figure shows that the
radial \taucl~is significantly higher for \pv~than for \ha~and,
moreover, that the linear dependence on the density for resonance
lines (as opposed to the quadratic dependence of recombination lines)
causes clumps to remain optically thick in \pv~throughout almost the
entire wind. The figure also illustrates how tangential \ha~photons
have higher optical depths than radial ones in the line forming
regions. Since non-radial photons are important for recombination
lines (Sect.~\ref{ha}), this enhances the effects from optically thick
clumping on the \ha~line formation in our empirical \lcep~model (as
seen in Fig.~\ref{Fig:stoch_prof}). The differences in clump optical
depths for radial ($\mu=1$) and tangential ($\mu=0$) photons stem from
the dependence $\tau_{\rm cl} \propto [d \varv /dr \mu^2 + \varv /r
(1-\mu^2)]^{-1}$, and the fact that $\varv /r$ is significantly lower
than $d \varv /dr$ in the relevant wind regions. In any case, based on
these simple estimates, one might expect that the basic results of
Sect.~\ref{theor} should hold in diagnostic applications of typical O
stars. That is, \ha~should be affected by optically thick clumping
only in the line core, whereas resonance lines should be much more
affected over the entire line profile.

\subsection{Constraints from inhomogeneous radiation-hydrodynamic models}

\begin{figure}
  \resizebox{\hsize}{!}{\includegraphics[angle=90]{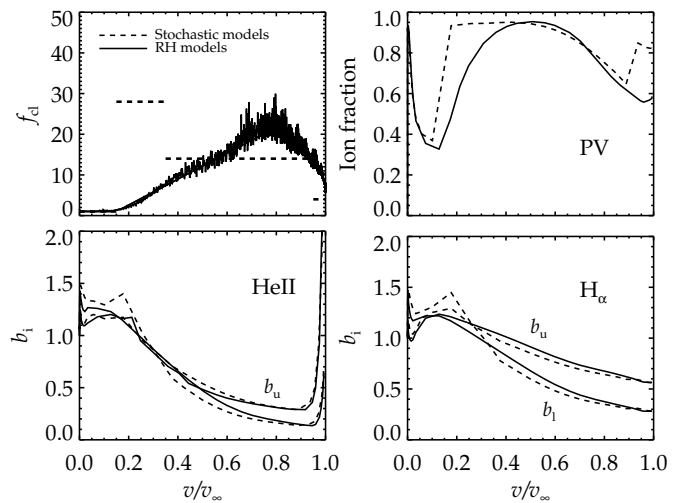}} 
  \caption{Clumping factors (\textit{upper left}), \pv~ionization
  fractions (\textit{upper right}), and \heii~and \ha~departure
  coefficients $b_{\rm i}$ (\textit{lower left} and \textit{lower
  right}, respectively) used in the RH and stochastic models
  of \lcep. Mean wind velocities are on all abscissas.}
\label{Fig:fcl}
\end{figure}

\begin{figure*}
  \resizebox{\hsize}{!}{\includegraphics[angle=90]{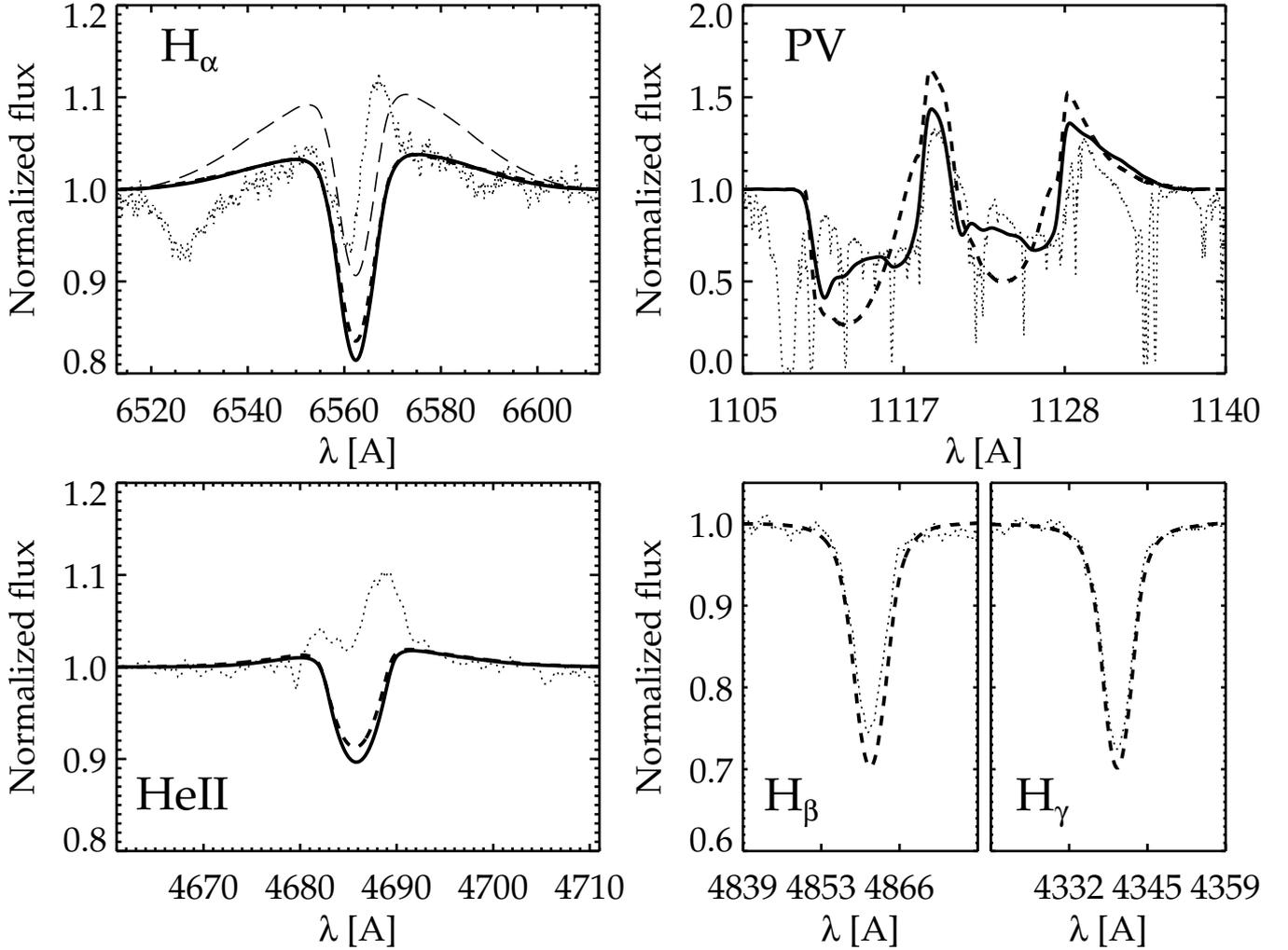}} 
\caption{Observed and synthetic line profiles for \lcep. \textit{Dotted lines} 
are the observations. \textit{Solid line} profiles are calculated from
the inhomogeneous radiation-hydrodynamic model
of \lcep~(Table~\ref{Tab:rh_par}), and \textit{dashed lines} from a
corresponding {\sc fastwind} model including microclumping.
The long-dashed line in the upper left panel is from a RH model in
which the density has been scaled to mimic an increase in the
mass-loss rate by 50\,\%.}
  \label{Fig:rh_prof}
\end{figure*}

\begin{figure*}
  \resizebox{\hsize}{!}
            {\includegraphics[angle=90]{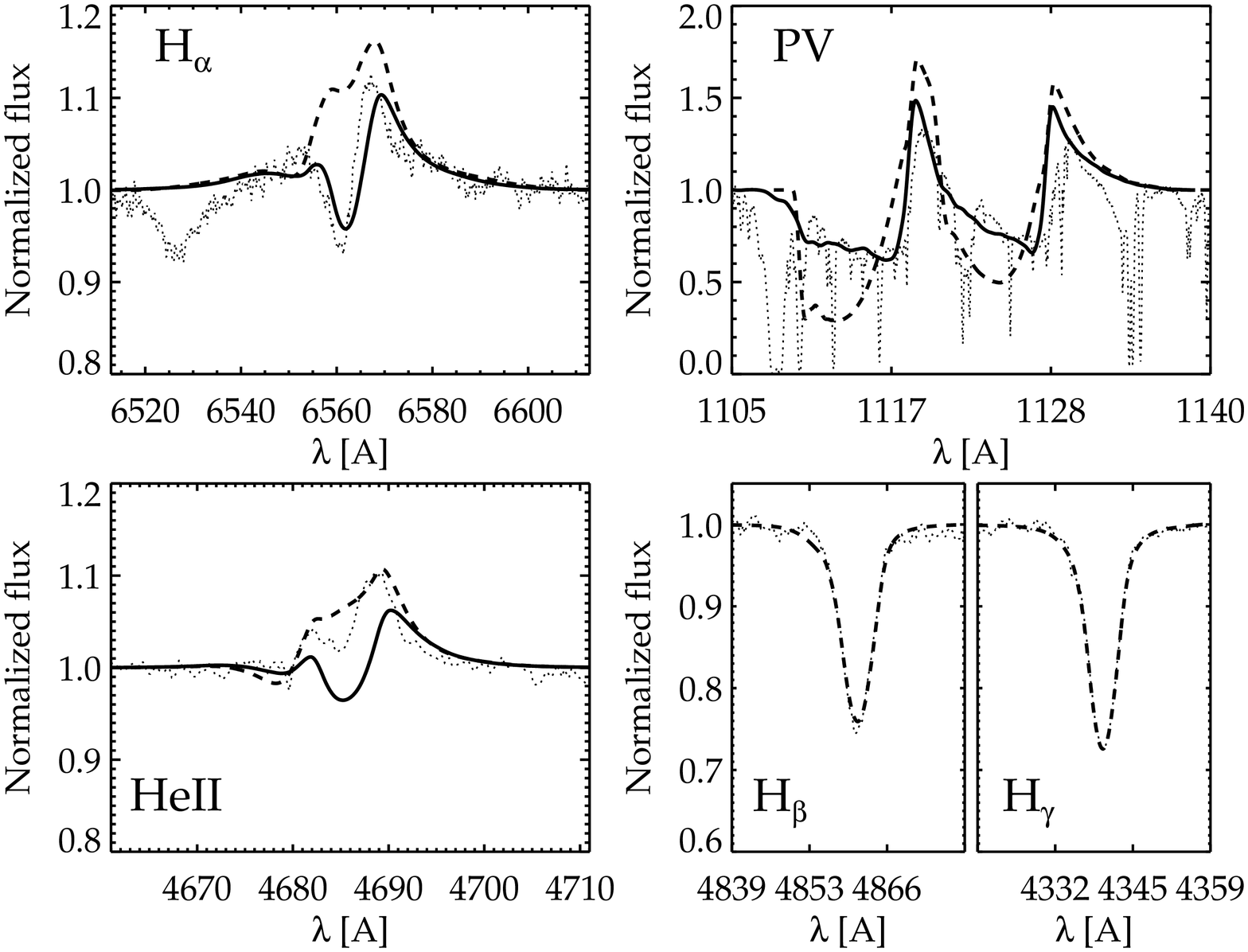}} 
            \caption{As Fig.~\ref{Fig:rh_prof}, but using our
            stochastic models (solids) with corresponding inferred
            empirical structure parameters (see
            Table~\ref{Tab:lcep_stoch} and text). The assumed
            mass-loss rate is the same as for the RH model of \lcep,
            see Table~\ref{Tab:rh_par}.}
\label{Fig:stoch_prof}
\end{figure*}

Fig.~\ref{Fig:rh_prof} displays line profiles calculated from our RH
model of $\lambda$ Cep. Consistent fits of the observed diagnostics
are not achieved. The \ha~line wings are reasonably well reproduced
but the core emission is much too low. The \pv~profiles are, actually,
better reproduced, although stronger than observed toward the blue
edge of the line complex (the `blue edge absorption dip' problem, see
Sect.~\ref{analytic}). The reasonable \pv~fits are due both to
adopting a rather low mass-loss rate for \lcep~(see
Table~\ref{Tab:rh_par}) and to lower velocity spans in these RH models
than in those analyzed in \one\footnote{The exact reasons for the
lower spans are still under investigation, and will be reported in a
future paper.}. The mass-loss rate was essentially chosen from a best
compromise when considering the complete diagnostic set, however note
that a consistent fit to all diagnostics could not be achieved,
independent of which mass-loss rate was adopted, as now discussed.

The apparent mismatch between \ha~emission in the core and in the
wings occurs because \fcl~increases rather slowly with increasing
velocity (Fig.~\ref{Fig:fcl}), which for a given mass-loss rate
implies that the optical depths in the \ha~core forming regions are
too low as compared to the optical depths in the wing forming
regions.  \heii~is subject to the same mismatch as \ha, and also the
cores of $\rm H_\beta$ and $\rm H_\gamma$ are deeper than observed.
The latter feature occurs because the photospheric absorption profiles
are not sufficiently re-filled by emission from the only weakly
clumped inner wind. Thus, the optical wind diagnostics all indicate
that the clumping factor as a function of velocity in \lcep~differs
from that predicted by the RH simulations \citep[see
also][]{Puls06,Bouret08}. On the other hand, any significant increase
in the mass-loss rate to obtain a better fit of the higher Balmer
lines and the core of \ha~would produce stronger than
observed \ha~and \heii~line wings (as illustrated for \ha~in
Fig.~\ref{Fig:rh_prof}) and, vice versa, a reduction of the mass-loss
rate to obtain a better fit of the (blue edge of the) PV lines would
produce too weak wings.

\paragraph{Comparison with the microclumping technique.} We now 
compare results from above with those from a microclumped {\sc
fastwind} model having the same (smoothed) clumping factors as the RH
model. The \pv~profiles calculated using the {\sc fastwind} model are
stronger than those calculated using the RH model. We may characterize
this difference by the difference in the equivalent widths \wl~of the
absorption parts of the profiles. \wl~is roughly 15\,\% lower for the
RH model \citep[see also][]{Owocki08}. However, this moderate
reduction in profile strength actually corresponds to a reduction in
the mass-loss rate by a factor of approximately two, because of the
resonance lines' slow response to mass loss.

Resonance line profiles stemming from the RH and microclumping models
also display different line \textit{shapes}.  For RH models,
significant velocity overlaps stemming from the non-monotonic velocity
field ensure that the observed flux at the blue side of the line
center is accurately reproduced without invoking any artificial and
highly supersonic `microturbulence', as must be done when using smooth
as well as microclumping wind models. Although not analyzed here, also
the absorption at velocities $>$\vinf~of saturated resonance lines may
be reproduced by RH models without invoking additional
microturbulence \citep[][\one]{Puls93}. For \ha~and \heii, the RH and
microclumping models yield almost identical results. This occurs
because clumps are optically thin in these diagnostics throughout
almost the entire wind, due to the slow increase of \fcl~with mean
wind velocity, which in turn results in wind densities in the inner
wind unable to produce optically thick clumps (compare to the
empirical models in the following section).

\subsection{Constraints from empirical stochastic models}
\label{stoch}

Clearly, the RH models fail to deliver satisfactory line profiles when
their structures are confronted with UV and optical wind diagnostics
in parallel. Here we use our stochastic models to modify the
wind-structure parameters and show how the results then may be
reconciled. This is a first attempt toward our long-term aim of using
consistent multi-diagnostic studies to obtain unique views
of \textit{empirical} mass-loss rates and structure properties of hot
star winds.

The mass-loss rate is determined by a best fit to the complete
diagnostic set (giving highest weight to the optical hydrogen
lines). We derive the same rate as was previously adopted for the RH
model of \lcep. This rate (\mdot=1.5\,$\times$\,\mdotun) is
approximately two times lower than the corresponding theoretical one
obtained using the mass-loss recipe in \citet{Vink00}
(\mdot=3.2\,$\times$\,\mdotun). In the outermost wind, we for now
adhere to the constraints on \fcl~derived from radio emission
by \citet{Puls06}, scaled with respect to the mass-loss rate derived
here. In the inner wind, both the distinct shape of \ha~in
$\lambda$~Cep\footnote{which only resembles the P Cygni shapes of the
UV resonance lines, since it is formed differently.} and the cores of
the higher Balmer lines may be used as tracers of structure.
The \ha~absorption trough followed by the steep incline to rather
strong emission can only be reproduced by our models if clumping is
assumed to start quite late \citep[see also][]{Puls06,Bouret08}, at a
velocity marginally lower than predicted by the RH models, however
with a much steeper increase with velocity (see
Fig.~\ref{Fig:fcl}). Also, in the particular case of \lcep, the upper
limit of the mass-loss rate derived by \citet{Puls06}
(\mdot=3.0\,$\times$\,\mdotun, inferred by assuming a smooth outermost
radio emitting wind) results in densities so high in the lowermost
wind that the \ha~trough never reaches below the continuum flux.
Moreover, additional constraints come from the cores of the higher
Balmer lines; the higher the densities in the lowermost wind, the
stronger the re-filling of the photospheric absorption profile by wind
emission.  Here as well the upper limit from \citeauthor{Puls06}
provides shallower than observed line cores. Thus, if we
require \fcl=1 at the wind base, and if our interpretation of the
abrupt shift from absorption to emission in \ha~as due to clumping is
correct, rather tight constraints on the mass-loss rate may be
obtained using only optical diagnostics.

The \ha~time series of \citet{Markova05} reveal that both the height
of the emission peak and the depth of the absorption trough depend on
the observational snapshot, variations can reach 0.04 in residual flux
units. Therefore it is not critical that neither the peak nor the
trough is perfectly reproduced by our models in
Fig.\ref{Fig:stoch_prof} (which displays a `representative'
observational snapshot). On the other hand, the observations do not
indicate any significant variation in the \textit{position} of the
emission peak. This might be an issue, because the late onset of
clumping redshifts the emission peak too much (at least when
neglecting differential rotation, see Sect.~\ref{m_est}), whereas an
earlier onset of clumping fails to produce an absorption trough. The
offset in the position of the emission peak is larger than the
estimated uncertainty in the radial velocity correction, which may
indicate that clumping is only partly responsible for the shape of the
\ha~core. Indeed, other interpretations have been suggested, 
and we comment on this in Sect.~\ref{m_est}.

The line shape of \heii~is well reproduced by our stochastic models,
but not the emission strength. The line reacts similarly to clumping
as \ha. In order to increase the central emission to the observed
level we would have to raise the clumping factor in the inner wind
even more, which in turn would produce stronger than
observed \ha~emission as well as shallower than observed $\rm H_\beta$
and $\rm H_\gamma$ cores.  Since hydrogen generally has more reliable
and robust departure coefficients than helium, we have given higher
weights to fits of hydrogen lines.  Interestingly, He\,II\,4686 shows
a similar offset as \ha~in the position of the emission peak.
  
The PV resonance lines are much more sensitive to the wind structure
parameters (see Sect.~\ref{analytic}) than to the mass-loss
rate. Hence these lines should be used only as a consistency check of
mass-loss rates derived from other diagnostics. Using the structure
parameters given in Table~\ref{Tab:stoch_par}, our stochastic models
yield reasonable fits of the PV lines. We use values
of \dt~and \xic~as in \one, including a higher \dt~in the outer wind
to account for clump-clump collisions, but are able to adopt a higher
value of $|$\dv/\dvb$|$, which however is still lower than predicted
by the RH models. This higher value stems from that we here consider
also optical diagnostics and from these derive a lower mass-loss rate
and higher clumping factors than what was assumed in Paper\,I,
essentially meaning that larger velocity spans then can be used when
fitting the \pv~lines.
 
\fcl~in the inner wind is
drastically different from that predicted by our RH model
for \lcep~(Fig.\ref{Fig:fcl}), and indicates that present-day RH
simulations fail to predict observationally inferred clumping factors,
at least for the inner wind. Regarding the outermost wind, let us
point out that the RH simulations used here only extend to $r \approx
35$, at which \fcl~is still decreasing. Simulations
by \citet{Runacres02}, which extend to much larger radii, indicate
that the clumping factor settles at $\approx 4$ in the outermost
wind. $f_{\rm cl} \approx 4$ is consistent with our derived mass-loss
rate and the constraints from radio emission derived by \citet{Puls06}
(see above). This suggests that the outermost wind is better simulated
by current RH models than the inner one.
    
\paragraph{Comparison with the microclumping technique.}
Here we compare the stochastic models from above with microclumped
models calculated with the same clumping factors.  When using the
microclumping technique, the PV resonance lines are not
\textit{directly} affected by the structured wind. The mass-loss rate adopted
in the previous paragraph then produces much too strong absorption in
these lines, see Fig.~\ref{Fig:stoch_prof}. Moreover, the high
clumping factor in the inner wind adopted in our stochastic models
results in so high densities that the clumps become optically thick
in \ha~and \heii~as well. This generally leads to weaker emission for
the stochastic models than for the microclumped ones (Sect.~\ref{ha}),
and \fcl's drastic increase from 1 to 28 makes the deviation from the
microclumping approximation prominent in this particular case. We have
confirmed that the same emission strength reduction results when using
our simplified analytic approach (Sect.~\ref{ha}), which supports the
rather strong emission reduction that we find in the \ha~core as well
as indicates that our analytic approach indeed might be a promising
tool for a consistent implementation into atmospheric NLTE codes.

In order to obtain reasonable fits of the PV lines within the
microclumping approximation we had to lower the mass-loss rate
significantly, to \mdot=0.4\,$\times$\,\mdotun~(this is the so-called
`\pv~problem', see also \citeauthor{Fullerton06}~2006). In turn this
meant that extreme clumping factors, $f_{\rm cl} \sim 400$, in the
inner wind were required to meet the observed amount of \ha~wind
emission. However, we have not been able to achieve a consistent fit
of the optical diagnostics using these highly microclumped {\sc
fastwind} models: if for example \ha~is fitted then the \heii~emission
is much too weak. Overall, the results in this section support the
view that the extremely low empirical mass-loss rates previously
indicated from \pv~might be a consequence of neglecting optically
think clumping when synthesizing resonance lines.

\section{Discussion}
\label{discussion}

\subsection{Are O star mass-loss rates reliable?}
\label{m_est}

\paragraph{Theoretical rates.} 
The time/spatial averaged mass-loss rate of our \lcep~RH model differs
from the rate of the corresponding smooth start model (used for
initialization) by less than 5\,\%. From this one might expect that
the clumped stellar wind should not significantly affect theoretical
mass-loss rates based on the line-driven wind theory.
However, \citet{Krticka08} (see also Muijres et al.~2010, submitted to
A\&A) made some first tests and included wind inhomogeneities in a
(steady-state) theoretical wind model of an O star. They found that
the predicted mass-loss rate increased when clumps were assumed to be
optically thin, because of increased recombination rates that shifted
the ionization balance to lower ionic states with more effective
driving lines. On the other hand, their tentative attempts to account
for optically thick clumps in the \textit{continuum} opacity as well
as for clumps with longer length scales than the Sobolev length
reduced the line force and led to lower predicted rates.

The reduced profile strengths of resonance lines (which are the main
drivers of the wind) found here should in principle also reduce the
line driving in theoretical steady-state wind models, but let us point
out that many lines that significantly contribute to the total driving
force might still be saturated because of the non-void interclump
medium. Nevertheless, it is clear that a thorough investigation of the
impact of clumping on predicted mass-loss rates is urgently
needed. The mass-loss rate for \lcep~derived here is approximately a
factor of two lower than the theoretical rate predicted by the
mass-loss recipe in \citet{Vink00}.

\paragraph{Empirical rates.} Our empirical mass-loss rate for 
\lcep~is 4.5 times lower than the rate inferred from synthesizing 
\ha~using a smooth wind model \citep{Repolust04}. The best constraints 
on the mass-loss rate in our analysis come from the distinct shape of
the \ha~line core and the higher Balmer lines (Sect.~\ref{stoch}).
Rotation in our models is treated by the standard convolution
procedure. But \lcep~is a fast rotator (Table~\ref{Tab:rh_par}), so
differential rotation might influence the formation of the line
profiles, particularly the \ha~core. Bouret et al. (published
in \citealt{Bresolin08}) found that the \ha~line in \zpup~ can be
fitted by assuming that clumping starts close to the wind
base, \textit{if} differential rotation is treated consistently.
Since \zpup~and \lcep~display similar \ha~profiles, it is possible
that the same effect could be at work also in the latter star, and
thereby that the rather late onset of and the rapid increase of
clumping in our stochastic model of \lcep~could be somewhat
exaggerated. Naturally, this could then also affect the inferred
mass-loss rate, since with a modified run of the clumping factor
another rate might be required to obtain a simultaneous fit of the
observed diagnostic lines.

The influence of X-ray and {\sc xuv/euv} radiation as created by
shocked wind regions \citep{Feldmeier97} on the occupation numbers is
not included in our analysis. These contributors are not important for
calculations of hydrogen occupation numbers \citep{Pauldrach01}, but
their significance for the ionization fractions of phosphorus is still
debated \citep{Krticka09,Waldron10}. We have used the alternative
unified atmospheric code {\sc wm}-Basic \citep{Pauldrach01}, which
treats X-ray and {\sc xuv/euv} radiation but \textit{not} wind
clumping, to estimate the impact of X-rays on the \pv~ionization
fractions. We find that effects are negligible at wind velocities
lower than $\varv/\varv_\infty \approx 0.5$ but profound at higher
velocities, with the \pv~ionization fraction significantly reduced
when X-rays (and of course the corresponding {\sc xuv/euv} radiation
tail) are included. This suggests that a proper treatment of these hot
radiation bands might resolve the earlier discussed `blue absorption
dip' problem, which is clearly visible in the \pv~line profiles
calculated from RH models (Fig.~\ref{Fig:rh_prof}, but note that we
overcame this problem in our stochastic models by increasing the
distances between clumps in the outermost wind, see
Table~\ref{Tab:lcep_stoch}).
           
\subsection{Structure properties of the clumped wind} 
\label{structure}

We identify two main problems when confronting synthetic
spectra from the time-dependent RH simulations of the line-driven
instability with observed lines in the UV and optical: i) the
absorption toward the blue edge of unsaturated UV resonance lines is
too deep in the simulations, and ii) the emission in the core of
\ha~is much too weak as compared to the emission in the wings. 
The first problem is related to the high predicted velocity spans in
the RH models, and was extensively discussed already in
Paper\,I. Moreover, in Sect.~\ref{m_est} we commented on that even if
the large velocity spans turn out to be stable features, this problem
might be overcome by a proper treatment of X-rays in the calculations
of ionization fractions.

The second problem arises because the predicted clumping factors in
the inner wind are too low as compared to those in the outer wind
(Fig.~\ref{Fig:fcl}). However, let us point out that velocity as well
as density perturbations in the inner wind of our RH simulation may be
overly damped, because we use the so-called smooth source function
(SSF) approximation when calculating the contribution to the line
force from the diffuse, scattered radiation field. In simulations that
relax the SSF approximation and account for gradients in the perturbed
source function (via an `escape-integral source function' formulation,
EISF, \citealt{Owocki96,Owocki99}), the structure in the inner wind is
more pronounced and also develops closer to the photosphere.

In any case, however, it is questionable if \textit{self-excited}
instability simulations will be able to reproduce the observed
clumping patterns (which have been found also in earlier
investigations based on the microclumping approximation,
e.g., \citealt{Bouret05,Puls06}), especially considering that our RH
model of \lcep~actually already is triggered (Table~\ref{Tab:rh_par}),
using Langevin perturbations mimicking photospheric
turbulence \citep{Feldmeier97}. Thus, while observations tracing the
outer wind seem to confirm the structures predicted by the line-driven
instability, observations tracing the inner wind might require the
consideration of an additional triggering mechanism to be reproduced,
which perhaps must be stronger than what is currently assumed. For
example, \citet{Cantiello09} proposed that gravity and/or acoustic
waves emitted in sub-surface convection zones may travel through the
radiative layer and induce clumping already at the wind base. However,
regarding gravity waves, it is not certain that these would have high
enough frequencies (i.e., higher than the atmosphere's acoustic cutoff
frequency) that they can be radially transported through the wind.
Another possibility for a strong clumping trigger might be non-radial
pulsations in the photosphere. Certainly it would be valuable to
investigate to what extent such triggers, within a line-driven
instability simulation using the EISF formulation, could produce
clumping patterns in the inner wind more compatible with the
observations.
 
\section{Additional considerations}
\label{future}

In this section, we discuss two applications for the analytic
formulation of line formation in clumpy winds presented in
Sect.~\ref{analytic}.

\subsection{Weak wind stars} 

The so-called weak wind problem is associated with observations of
(primarily) O-dwarfs of late types, which appear to have mass-loss
rates much lower than what is predicted by the line-driven wind
theory, and also much lower than other `normal' O stars of earlier
spectral types. However, a major problem with wind diagnostics in this
domain is that the primary optical diagnostic, \ha, becomes
insensitive to changes in the mass-loss rates, so that only upper
limits can be inferred from this line. Therefore one must for these
objects quite often rely solely on the intrinsically stronger UV
resonance lines. For a comprehensive discussion on the weak wind
problem, see \citet{Puls08}.

\begin{figure}
  \resizebox{\hsize}{!}{\includegraphics[angle=90]{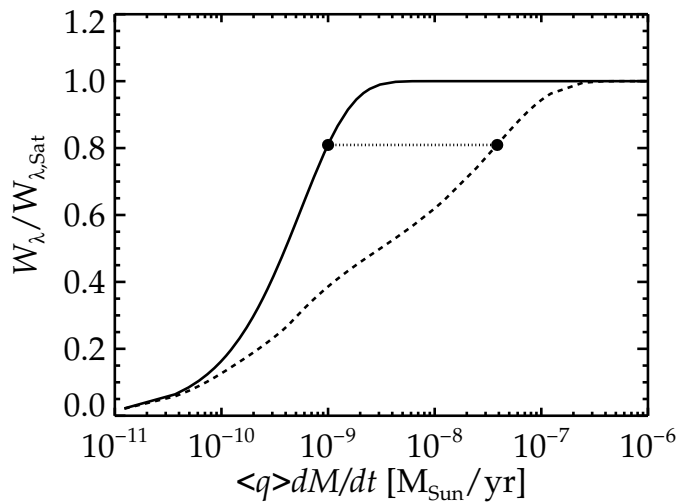}} 
  \caption{Equivalent widths, $W_\lambda$ (normalized to the value for
  a saturated line), for the absorption part of the N{\sc v} resonance
  line at 1240 $\AA$, as functions of the product of the ionization
  fraction of N{\sc v}, $\langle q \rangle$, and the mass-loss rate.
  The solid line is calculated from smooth models and the dashed line
  from structured ones. The black dots denote $W_\lambda$'s for models
  corresponding to a smooth model with $\langle q \rangle \dot{M} =
  10^{-9}$, see text.}
  \label{Fig:ewcurve}
\end{figure}

In the following, we demonstrate the potential impact of optically
thick clumping on diagnostic resonance lines in weak wind stars using
the analytic formulation developed in Sect.~\ref{analytic}. We use one
component of the N{\sc v} doublet at 1240\,$\AA$, assume a solar
nitrogen abundance \citep{Asplund05b}, and take a generic O-dwarf with
parameters \rstar=8.0\,\rstun, \vinf=1500\,\kms, and $Y_{\rm
He}=0.1$. Since in this section we only discuss predictions for the
product of mass-loss rate and ionization fraction $q$ for resonance
lines, no effective temperature needs to be specified (see
Eqs.~\ref{Eq:kappa0_tot}-\ref{Eq:tausob_app}).  The N{\sc v} doublet
was among the lines utilized in the study of \citet{Marcolino09}, and
also our chosen parameters correspond well to the parameters for the
five stars analyzed and found to have very weak winds (more than an
order of magnitude lower than predicted by theory) in that study. To
avoid problems regarding the onset of clumping and the aforementioned
`blue absorption dip', we consider only the velocity interval
$\varv/\varv_\infty = 0.25 -0.75$. Absorption-part line profiles for
structured winds are calculated using Eq.~\ref{Eq:Fow_1} and adopting
the same structure parameters as in Sect.~\ref{analytic}
(\fv=0.25, \xic=0.0025, \dt=0.5, and a smooth `$\beta$=1' velocity
field).

Fig.~\ref{Fig:ewcurve} shows the curve-of-growths for structured and
smooth models, respectively, as functions of the mean ionization
fraction of N{\sc v} times the mass-loss rate, $\langle
q \rangle \dot{M}$. Clearly, mass-loss rates derived from smooth
models may be severely underestimated also for stars with weak
winds. For example, if we for this star were to infer $\langle
q \rangle \dot{M}=10^{-9} \ \rm M_\odot/yr$ from a smooth model, the
corresponding rate inferred from a structured one would be $(\langle
q \rangle \dot{M})_{\rm struc}=3.8 \times 10^{-8} \ \rm M_\odot/yr =
38 (\langle q \rangle \dot{M})_{\rm smooth}$ (see
Fig.~\ref{Fig:ewcurve}). Thus, if using smooth models (or
microclumped, since microclumping has no effect on the resonance lines
as long as no significant changes occur in the ionization fractions),
one could easily derive mass-loss rates more than an order of
magnitude lower than corresponding rates derived from structured
models (see also \citealt{Oskinova07} and Paper\,I), and thereby one
could also misinterpret observations as suggesting that mass-loss
rates are much lower than predicted by theory.

We emphasize, however, that this simple example merely demonstrates
how optically thick clumping \textit{might} be important also for
resonance line diagnostics in so-called weak wind stars, and
that, \textit{if} the winds are clumped, one must be careful not to
simply assume that strongly de-saturated resonance lines also imply
optically thin clumps. The actual mass-loss reductions will depend
critically both on the assumed ionization fractions and on the adopted
structure parameters.  Thus, a multi-diagnostic study (to constrain
the structure parameters), including a detailed consideration of
X-rays (to obtain reliable ionization fractions), is required for more
quantitative results.  Nevertheless, we may safely say that, because
of these inherent problems in UV line diagnostics, it is important to
put further constraints on the weak wind problem by exploiting other
diagnostics that are sensitive to mass loss but neither have optically
thick clumps nor are affected by X-rays (as is probably true for,
e.g., the infrared Br$_\alpha$ line, Najarro et al.~2010, submitted
to A\&A, see also \citealt{Puls09b}).

\subsection{Resonance line doublets} 
\label{doublets}

\citet{Massa08} pointed out that additional empirical 
constraints on wind structure may be obtained by considering the
observed profile-strength ratios of resonance
line \textit{doublets}. The line strength parameter, $\kappa_0$, of
such doublets is in proportion to the oscillator strengths of the
individual components, $f$, which for the cases of interest here are
$f^{\rm b}/f^{\rm r}=2$, with superscripts $b$ and $r$ denoting the
blue and red line components, respectively. However, if clumps are
optically thick for the investigated lines, the resulting
profile-strength ratio may deviate quite significantly from the one
implied by smooth modeling (see discussion in Paper\,I).  For example,
in the case of very optically thick clumps and a void interclump
medium, Eq.~\ref{Eq:Fow_1} simply gives $R_{\rm a,x} = 1 - \xi_{\rm
x}$, i.e. the profile strength becomes independent of $\kappa_0$. The
analogy for continuum diagnostics, or for line diagnostics in
a \textit{non-accelerating} medium, is the well-known result that for
a medium consisting of infinitely dense absorbers embedded in a
vacuum, the effective opacity is independent of the atomic opacity
(see footnote 4 in Appendix A). Also for such a situation would the
inferred profile-strength ratio be exactly one.

A major advantage of this line diagnostic is that the dependence on
X-rays should cancel out. Recently, \citet{Prinja10} extended
the \citeauthor{Massa08} work to include a large number of B
supergiants, for which they, from the Si\,{\sc iv} $\lambda\lambda
1400$ resonance doublet, derived empirical line-strength ratios,
$\kappa_0^{\rm b}/\kappa_0^{\rm r}$, using smooth wind models. The
stars showed a wide spread between unity and the predicted factor of
two, with the majority of them lying in the range 1.0 to 1.5, and with
an overall mean of 1.46 (standard deviation $\sim$0.31). In the
following, we shall discuss this diagnostic under the assumption that
the doublet components are well separated, so that each component can
be treated as a single line, which is reasonable for, e.g., the just
mentioned silicon lines in typical B-supergiants and for \pv~in
OB-stars.

We now show that our analytic formulation for resonance line formation
indeed predicts profile-strength ratios on the same order as those
discussed above.  Following the preceding section, we assume a solar
abundance for silicon, make use of a generic B-supergiant
with \rstar=30.0\,\rstun, \vinf=800\,\kms, and $Y_{\rm He}=0.1$, adopt
the same structure parameters as in the previous section, and consider
only the velocity interval $\varv/\varv_\infty = 0.25-0.75$. We then
assume that for this generic star we derive $\langle q \rangle \dot{M}
= 5 \times 10^{-9}$ from the Si\,{\sc iv} resonance doublet formed in
a \textit{structured} wind model. By once more exploiting the
curve-of-growth (as in Fig.~\ref{Fig:ewcurve}, but now for the two
components of Si\,{\sc iv}), we can then easily translate the
structured results to corresponding smooth ones. We find a ratio
$(\kappa_0^{\rm b}/\kappa_0^{\rm r})_{\rm smooth} \approx 1.4$, which
agrees well with the results derived by \citet{Prinja10}.

The doublet ratios are, in fact, almost ideal diagnostics regarding
structure properties, since all other dependencies cancel out.
Therefore ratios deviating from two might be the cleanest indirect
signatures of optically thick clumping that we presently have, and may
in principle be used to extract empirical information on the behavior
of $\xi$. We write the ratio of the blue and red absorption-part line
profile at frequency $x$ as
\begin{equation} 
	\frac{R^{\rm b}_{\rm a,x}}{R^{\rm r}_{\rm a,x}} = \frac{(1-\xi_{\rm x})e^{-(2\tau_{\rm ic,x}^{\rm r})} + \xi_{\rm x} e^{-(2\tau_{\rm cl,x}^{\rm r})}}
	{(1-\xi_{\rm x})e^{-\tau_{\rm ic,x}^{\rm r}} + \xi_{\rm x} e^{-\tau_{\rm cl,x}^{\rm r}}}.
	\label{Eq:br_ratio}
\end{equation} 
Generally, this equation can be solved for $\xi_{\rm x}$ only if the
line optical depths and the interclump densities are known (the
latter for example from observations of saturated resonance lines, see
Paper\,I). However, under certain circumstances we can eliminate the
need for such external knowledge. For example, assuming that
\textit{all clumps are optically thick}, we may write
\begin{equation} 
	  \frac{R^{\rm b}_{\rm a,x}}{R^{\rm r}_{\rm a,x}} = e^{-\tau_{\rm
		ic,x}^{\rm r}} = \frac{R^{\rm r}_{\rm
		a,x}}{1-\xi_{\rm x}} \ \ \rightarrow \ \ \xi_{\rm x} = 1
		- \frac{(R_{\rm a,x}^{\rm r})^2}{R_{\rm a,x}^{\rm
		b}}.  \label{Eq:brratio_2}
\end{equation}    
Applying the last expression to our line profiles computed for Si\,{\sc
iv} using Eq.~\ref{Eq:Fow_1} reveals a mean value of $\xi = 0.48$ in a
velocity bin $\varv/\varv_\infty = 0.4-0.5$, which agrees well with the actual
mean (calculated from the assumed structure parameters), $\xi = 0.51$.
Thus, this approximation can provide a quite good direct empirical
mapping of $\xi$, without any knowledge about optical depths etc.
Another case for which the profile-strength ratio can be directly
related to $\xi$ is that of a completely transparent background medium
(i.e., a void interclump medium in our case). That limiting case of
Eq.~\ref{Eq:br_ratio} has been long recognized and used by the quasar
community \citep[e.g.,][]{Ganguly99}, for the formation of intrinsic,
narrow absorption-line doublets.

However, let us point out that this theoretical example only
demonstrates that our basic formalism appears reasonable. In a real
application, there will be a contribution also from
the \textit{re-emission} part of the line profile, i.e., what we
actually measure from an observation is the total line profile $R_{\rm
x} = R_{\rm a,x} + R_{\rm em,x}$. Thus, to empirically infer $\xi_{\rm
x}$ from Eq.~\ref{Eq:brratio_2} (which involves $R_{\rm a,x} = R_{\rm
x} - R_{\rm em,x}$), we must either simply neglect the re-emission
contribution (which generally will not be possible) or actually
calculate $R_{\rm em,x}$, as predicted by a \textit{structured} wind
model. For resonance lines (as opposed to recombination lines, see
Sect.~\ref{ha}), a simplified approach for $R_{\rm em,x}$ in clumpy
winds is still to be developed; it is a very demanding task because of
the source function's scattering nature.  In principle though, a
treatment corresponding to the `smooth source function' formalism used
in our time-dependent RH simulations (see Sect.~\ref{structure}) might
be a reasonable first approximation.

\section{Summary and future work}
\label{summary}

We investigate diagnostic features for deriving mass-loss rates from
the clumped winds of hot, massive stars, without relying on the
microclumping approximation. It is found that present-day RH
simulations of the line-driven instability are not able to
consistently fit the UV and optical diagnostics in a prototypical
O-supergiant. By creating empirical stochastic wind models, we achieve
consistent fits mainly by increasing the clumping in the inner wind. A
mass-loss rate is derived that is approximately a factor of two lower
than what is predicted by theory. The best constraints come from the
optical diagnostics. The UV resonance lines are much more sensitive to
the wind's structure parameters (i.e., to the clumping factor, the
interclump medium density, etc.) than to the mass-loss rate, and
should, thus, not be the preferred choice when deriving empirical
mass-loss rates.

We discuss both recombination line and resonance line formation in
detail. Resonance lines always suffer the effects of optically thick
clumping in typical diagnostic lines, and their profiles are thereby
weaker for models with a detailed treatment of clumping than for
models that rely on the microclumping approximation. Recombination
lines are less affected because of the lower optical depths in typical
diagnostic lines. However, emission strength reductions as compared to
microclumped models are significant for stars with high mass-loss
rates (e.g., Wolf-Rayet stars) and can be so for O stars as
well, \textit{if}, for example, strong clumping is present in the
lower wind, as illustrated by our diagnostic study of \lcep.

An analytic method to model these lines in clumpy winds, without any
restriction to microclumping, is suggested and shown to yield results
consistent with those from detailed stochastic models. Some first
results are given, illustrating the potential significance of
optically thick clumps for diagnostic lines in weak wind stars, and
confirming recent results that profile-strength ratios of resonance
line doublets may be used as tracers of wind structure and optically
thick clumping. We intend to refine this method and incorporate it
into suitable NLTE unified atmospheric codes, in order to investigate
effects of optically thick clumping on the occupation numbers.

It is pivotal that 3D, time-dependent RH models of the line-driven
instability be developed, with an adequate treatment of the 3D
radiation transport. New models are required
to investigate whether the structure predicted by present-day
simulations is stable or a consequence of current physical assumptions
and simplifications.

\begin{acknowledgements}
{We thank the anonymous referee for detailed comments and
  suggestions. J.O.S gratefully acknowledges a grant from the
  International Max-Planck Research School of Astrophysics (IMPRS),
  Garching, and also current financial support from the DFG cluster of
  excellence.}
\end{acknowledgements}

\bibliographystyle{aa}

\appendix

\section{Analytic treatment of line formation in clumped hot star winds}
\label{appA}

\paragraph{Resonance lines.} We propose to write the absorption part 
of a resonance line formed (from a radial ray) in a clumped wind as

\begin{equation}
  R_{\rm a} = \xi e^{-\tau_{\rm cl}} + (1-\xi) e^{ -\tau_{\rm ic}},
  \label{Eq:Fow}
\end{equation} 

\noindent where $\xi$ is defined as the \textit{fraction of the velocity field 
over which photons may be absorbed by clumps}, the optical depths are
those for the clumped (subscript cl) and rarefied (subscript ic)
medium, and dependencies on the normalized, dimensionless frequency
$x$ have been suppressed for simplicity (cf. Sect.~\ref{analytic} in
main paper).

Following \citet{Owocki08} we define the \textit{velocity filling
factor} \fvel~as the fraction of the velocity field covered by clumps
(in full analogy with the volume filling factor \fv). That
is, \fvel~is the ratio of the velocity span of the clump, \dv, to the
velocity separation between two clump centers, \Dv,

\begin{equation}
	f_{\rm vel} \equiv \frac{\delta \varv}{\Delta \varv}.
\end{equation}	

\noindent In our stochastic models we have the clump velocity 
span $\delta \varv \approx |\delta \varv/\delta \varv_\beta|
(d \varv_\beta/dr) \delta r$ and from the definition of \fv~(see \one)
$\delta r \approx f_{\rm v}
\Delta r$, with $\Delta r = \varv_\beta \delta t$ 
the radial distance between two clump centers.  Similarly one may
approximate $\Delta \varv \approx (d \varv_\beta/dr) \Delta r$, which leads to

\begin{equation}
	f_{\rm vel} \approx |\frac{\delta \varv}{\delta \varv_\beta}| f_{\rm v}.
        \label{Eq:fvelex}
\end{equation} 

\noindent Thus, a smooth velocity law (\dv=\dvb) implies \fvel=\fv. 
The absolute value sign becomes important in any compressive wind
region, since optical depths must always be positive (see below).

Actually, Eq.~\ref{Eq:Fow} is in form equivalent to the analytic
transfer solution derived by \citet{Levermore86}, for the ensemble
averaged intensity in a two-phase $[{i=A,B}]$ Markovian model of a
static purely continuum absorbing medium in the limit that the
characteristic length scales $l_{\rm i}$ of the fluid packets of both
components are much longer than the domain of integration\footnote{We
mention in passing that the Levermore et al. model also yields the
result $e^{-s/h}$, with $s$ the path length, for the normalized
intensity in the limit of infinitely dense absorbers in a background
vacuum. This is equivalent to the result for a fully porous wind
obtained by, e.g., \citet{Owocki04}.}, if we just substitute $l_{\rm
A},l_{\rm B} \rightarrow \delta \varv, \Delta \varv$.  Thus, from this
analogy it is clear that we may set $\xi=f_{\rm vel}$ as long as the
Sobolev-like requirement $\delta \varv >> C \varv_{\rm t}$ is
satisfied, where $C \varv_{\rm t}$ is the velocity extent over which a
photon of frequency $x$ may be absorbed (that is, the velocity extent
of a resonance zone). This limiting situation corresponds to the case
that the line profile can be represented by a delta function, so that
the sharp edges of the resonance zones prevent any absorption at
frequencies not Doppler shifted to the very line center, resulting in
a localized radiative transfer. The optical depths in Eq.~\ref{Eq:Fow}
are then understood to be the Sobolev ones. That is, $\tau_{\rm cl}
= \tau_{\rm sm} / f_{\rm vel}$ and $\tau_{\rm ic} = \tau_{\rm sm}
(x_{\rm ic}/ f_{\rm vel})$, with $\tau_{\rm sm}$ the optical depth in
the smooth case (Eq.~\ref{Eq:tausob_app}).

However, especially in the outer wind (but, depending on the onset of
clumping, also in the innermost wind, see Fig.~\ref{Fig:analytic}) we
will generally have $\delta \varv < C \varv_{\rm t}$ and the effective
fraction of the velocity field over which photons can be absorbed by
clumps will increase. The exact form of the radiation transport is
then likely to be very complex. Nonetheless, let us in a first attempt
try to simply modify $\xi$ in order to account for the essential
effects. We write

\begin{equation}
	\xi \approx \frac{\delta \varv + C \varv_{\rm t}}{\Delta \varv},  
\end{equation} 

\noindent where the factor $C \varv_{\rm t}$ now 
represents a sort of correction to the limiting case of $\delta \varv
>> C \varv_{\rm t}$. A linear addition is chosen because the basic equation
determining whether or not a photon actually can be absorbed (i.e.,
whether or not it is located within its resonance zone) is $x_{\rm
cmf} = x_{\rm obs} - \varv$, with $x_{\rm cmf}$ and $x_{\rm obs}$ the
co-moving and observer's frame frequencies, respectively. 

The factor $C$ accounts for the fact that the `effective resonance
zone' over which photons can be absorbed by clumps is larger than that
provided by $\varv_{\rm t}$ (at least for relatively strong lines).
Photon absorption at $x$ within clumps is given by the distribution
function $e^{-\tau_{\rm cl,x}}$, with expectation value $\tau_{\rm
cl,x}=1$. Therefore we may estimate $C$ using the `effective profile
width', determined by solving for the co-moving frame frequency at
which unity optical depth is reached, \textit{if} a clump is present,

\begin{equation}
        \tau_{\rm cl} \frac{1-\rm erf[x_{\rm cmf}/\varv_{\rm t}]}{2} = 1,
        \label{Eq:unity}
\end{equation}

\noindent where erf is the error function. The effective profile width
then is $C=2x_{\rm cmf}/ \varv_{\rm t}$, where $x_{\rm cmf}$ is given
by the solution to Eq.~\ref{Eq:unity}. Note that $C$ now is allowed to
be velocity dependent, $C \rightarrow C(\varv)$.  In addition, the
expression for the clump optical depth should now be modified,
$\tau_{\rm sm}/f_{\rm vel} \rightarrow \tau_{\rm sm}/\xi$, to account
for the fact that individual clumps no longer cover a complete
resonance zone. 

With $C$ determined we can cast $\xi$ in the convenient form

\begin{equation}
	\xi \approx f_{\rm vel} + C \eta,  
\end{equation} 

\noindent where $\eta \equiv \varv_{\rm t}/\Delta \varv$ is 
the effective escape ratio.  Note the difference between this
definition of $\eta$ and that given in \one. The two are related as

\begin{equation}
	\eta = (1-f_{\rm vel}) / \eta_{\rm old},
\end{equation} 

\noindent where $\eta_{\rm old}$ denotes our earlier definition. The advantage of 
re-defining $\eta$ is that we may now separate out the porosity
dependence in $\xi$, writing

\begin{equation} 
\eta = \frac{\varv_{\rm t}}{\Delta \varv} \approx \frac{\varv_{\rm t}/(d \varv_\beta/dr)}{\Delta r} = \frac{L_{\rm r}}{h}, 
\end{equation}

\noindent with $h=\delta r/f_{\rm v} = \Delta r = \varv_\beta \delta t$ the
porosity length of the medium in our geometry and $L_{\rm r}
= \varv_{\rm t}/(d \varv_\beta/dr)$ the radial Sobolev length for a
smooth velocity field. The coupling between vorosity and porosity
becomes clear via $\eta$.

As defined, $\eta$ may in principle take arbitrarily high values, so
for the examples in this paper we simply set $\xi=1$ whenever $\xi \ge
1$, because in a wind with a smooth velocity field the clumps
obviously cannot absorb photons over a velocity space larger than that
covered by the $\beta$ velocity law. On the other hand, if we allow
for clumps to be randomly positioned in velocity space, overlapping
velocity spans will lead to a change in the effective coverage
fractions. If velocity perturbations are sufficiently large, one may
simply substitute $\xi \rightarrow (1-e^{-\xi})$ and permit $\xi$ to
take arbitrarily high values.  However, it is clear neither if
velocity perturbations will be sufficiently large nor how to handle
the case when more than one clump is crossed within a resonance
zone. Thus we for now consider only the simple case of a smooth
velocity field, deferring to future work a careful study of these
randomization effects. In any case, we note that our formalism
recovers the smooth optical depth $\tau_{\rm sm}$ in the limit
$\Delta \varv << C \varv_{\rm t}$ (as expected because then the
individual clumps obviously are optically thin).

Finally, Eq.~\ref{Eq:Fow} has the proper behavior in the limiting
cases of a smooth or microclumped wind. For the former (\xic=1
and \taucl=$\tau_{\rm sm}$),

\begin{equation}
  R_{\rm a} = e^{-\tau_{\rm sm}},
\end{equation} 
 
\noindent and for the latter ($\tau_{\rm cl} << 1$) some simple algebra yields, 

\begin{equation}
  R_{\rm a} \approx 1-\tau_{\rm sm},  
\end{equation} 
 
\noindent where we recall that this last result is expected because 
resonance line formation depends linearly on the density (see
Sect.~\ref{stoch_par}).

\paragraph{Recombination lines.} The absorption part of recombination lines 
such as \ha~may also be approximated as described above. Furthermore,
since the source function in these lines can be prescribed (see
Sect.~\ref{models}), we can make a similar approximation for the
re-emission part

\begin{equation}
  R_{\rm em} = S \xi (1 - e^{-\tau_{\rm cl}}) + S(1-\xi)(1 - e^{ -\tau_{\rm ic}}),
  \label{Eq:reem}
\end{equation} 

\noindent where $S$ is the source function at the resonance point in units of
the continuum intensity. The total line profile $R_{\rm x}$ is then
given by $R_{\rm x}=R_{\rm a,x}+R_{\rm em,x}$.  It is important to
realize that the re-emission profile is much more influenced by
non-radial photons than is the absorption part profile. Thus we
replace the radial approximation for $\xi$ with a corresponding
line-of-sight expression, $\xi \rightarrow \xi_{\rm z}$, by
substituting $L_{\rm r} \rightarrow L_{\rm z}$ and $h \rightarrow
h/\mu$, where curvature effects for a clump have been neglected, and
now obtain the final line profiles by performing the standard
integration over a pre-specified number of P-rays. The optical depths
from the previous paragraph must be replaced by corresponding ones for
recombination lines (see Sect.~\ref{ha}), where of course care must be
taken for the now angular dependent $\tau$.

Also Eq.~\ref{Eq:reem} has the proper behavior for smooth as well as
for microclumped winds.  In the same manner as for the resonance lines,
we obtain for the former

\begin{equation}
  R_{\rm em} = S (1 - e^{-\tau_{\rm sm}}), 
\end{equation} 

\noindent and for the latter 

\begin{equation}
  R_{\rm em} \approx S \tau_{\rm sm} f_{\rm cl},  
\end{equation} 

\noindent which is expected because recombination line formation depends 
on the square of the density (see Sect.~\ref{stoch_par}).

Comparisons between the analytic approximations outlined here and
numerical simulations using our stochastic wind models and detailed
radiative transfer codes are given in the main paper.


\end{document}